\documentclass[journal]{IEEEtran}

\usepackage{subfigure}
\usepackage{amsmath,amssymb}
\usepackage[dvips]{graphicx}
\usepackage{amsfonts}
\usepackage[mathscr]{eucal}
\usepackage{latexsym}
\usepackage{amsthm}
\usepackage{exscale}
\usepackage[mathscr]{eucal}
\usepackage{bm}
\usepackage[dvipsnames]{color}
\usepackage{cases}
\usepackage{epsfig}
\usepackage[center,small]{caption}
\usepackage{algorithm}
\usepackage{algorithmic}
\usepackage[verbose,nospace,sort]{cite}
\usepackage{tabularx}
\usepackage{multirow}
\usepackage{multicol}
\usepackage{balance}

\graphicspath{{./figsJournalNew/}}

\scrollmode

\newtheorem{theorem}{Theorem}

\newtheorem{lemma}{Lemma}

\newcommand{\LB}{\mathrm{lb}}

\newcommand{\Sn}{\mathrm{S}}
\newcommand{\D}{\mathrm{D}}
\newcommand{\R}{\mathrm{R}}
\newcommand{\SR}{\mathrm{sr}}
\newcommand{\RD}{\mathrm{rd}}

\newcommand{\opt}{*}
\newcommand{\Rcwf}{R^{\mathrm{cwf}}}
\newcommand{\pcwf}{p^{\mathrm{cwf}}}

\begin{document}
\title{Throughput Maximization for Mobile Relaying Systems}
\author{Yong~Zeng, Rui~Zhang, and Teng Joon Lim \\
\thanks{The authors are with the Department of Electrical and Computer Engineering, National University of Singapore. e-mail: \{elezeng, elezhang, eleltj\}@nus.edu.sg.}
\thanks{Part of this work has been submitted to IEEE Global Communications Conference (Globecom), 2016.}
\vspace{-4ex}
}

\maketitle

\begin{abstract}
Relaying is an effective technique to achieve reliable wireless connectivity in harsh communication environment. However, most of the existing relaying schemes are based on relays with fixed locations, or \emph{static relaying}. In this paper, we consider a novel \emph{mobile relaying} technique, where the relay nodes are assumed to be capable of moving at high speed. Compared to static relaying, mobile relaying offers a new degree of freedom for performance enhancement via careful relay trajectory design. We study the throughput maximization problem in mobile relaying systems by optimizing the source/relay transmit power along with the relay trajectory, subject to practical mobility constraints (on the relay speed and initial/final relay locations), as well as the \emph{information-causality constraint} at the relay owing to its decode-store-and-forward (DSF) strategy. It is shown that for fixed relay trajectory, the throughput-optimal source/relay power allocations over time follow a ``staircase'' water filling (WF) structure, with \emph{non-increasing} and \emph{non-decreasing} water levels at the source and relay, respectively. On the other hand, with given power allocations, the throughput can be further improved by optimizing the relay trajectory via successive convex optimization. An iterative algorithm is thus proposed to optimize the power allocations and relay trajectory alternately. Furthermore, for the special case with free initial and final relay locations, the jointly optimal power allocation and relay trajectory are derived. Numerical results show that by optimizing the trajectory of the relay and power allocations adaptive to its induced channel variation, mobile relaying is able to achieve significant throughput gains over the conventional static relaying.
\end{abstract}


\section{Introduction}
In wireless communication systems, relaying is an effective technique for throughput/reliability improvement as well as range extension, which has drawn significant interests over the past few decades \cite{650,23,24,634,639,640}.  However, due to the practical  constraints such as limited node mobility and wired backhauls, most of the existing relaying techniques are based on relays deployed  in fixed locations, or {\it static relaying}. In this paper, we study a new relaying technique, termed {\it mobile relaying}, where the relay nodes are assumed to be capable of moving at relatively high speed, e.g., enabled by terminals mounted on ground or aerial vehicles. We note that the practical deployment of high-mobility nodes dedicated for wireless relaying is becoming more feasible than ever before, thanks to the continuous cost reduction in autonomous or semi-autonomous vehicles, such as unmanned aerial vehicles (UAVs) \cite{616,618,649}, as well as the drastic device miniaturization in communication equipment.

 Compared with the conventional static relaying, mobile relaying has several promising advantages. First, on-demand mobile relaying systems are more cost-effective and can be much more swiftly deployed, which make them especially suitable for unexpected or limited-duration events \cite{615}, such as emergency response, military operation, etc. Besides, the high mobility of mobile relays offers new opportunities for performance enhancement through the dynamic adjustment of relay locations to best suit the communication environment, a technique that is especially promising for delay-tolerant applications \cite{653,652,655}, such as periodic sensing, large data uploading/downloading, etc. Note that while node mobility has been well exploited for upper layer designs in communication networks \cite{651,638,656}, its exploitation for more efficient physical layer designs is still under-developed.

To realize the full potential of mobile relaying techniques, we consider in this paper the classic three-node cooperative communication system consisting of fixed source and destination nodes assisted by a mobile relay.  We study the throughput maximization problem for this mobile relaying system by optimizing both the relay trajectory and the source/relay power allocations over a finite time horizon. Note that for mobile relaying systems, trajectory planning and adaptive communication are two important design aspects that are closely coupled with each other. On one hand, adaptive communication such as transmit power allocation should exploit the predictable channel variation induced by relay movement, e.g., the source/relay should transmit with more power when the relay moves closer to the source/destination to exploit better channels. On the other hand, the optimal relay trajectory design needs to strike a balance between the source-relay and relay-destination throughput, which also depends on the power allocation at the source/relay transmitters. To tackle such a tradeoff, we jointly optimize the transmit power allocations and relay trajectory to maximize the throughput, subject to the average transmit power constraints at the source/relay, as well as the practical mobility constraints on the relay maximum speed and its initial and final locations. Furthermore, unlike the conventional static relaying \cite{639},\cite{640}, we propose a new {\it decode-store-and-forward} (DSF) strategy for the mobile relay to maximally exploit the movement-induced channel variations. With DSF, the data received by the relay from the source is temporarily stored in a buffer, if necessary,  before being forwarded to the destination. We therefore need to consider the {\it information-causality constraint} at the relay, i.e., the relay can only forward the data that has been received from the source previously. Note that compared to conventional static relaying with essentially instantaneous information forwarding in the time scale of symbol or packet duration, information-causality constraint is more critical for the mobile relaying with DSF strategy, where the data may need to be buffered for much longer duration for the relay to reach a better position for information forwarding. Though a larger delay may have to be tolerated by some of the packets transmitted, mobile relaying with optimally designed DSF strategy is able to achieve significant throughput gains over the conventional static relaying, as will be shown in this paper. Specifically, the main contributions of this paper are summarized as follows.

\begin{figure}
\centering
\includegraphics[scale=0.65]{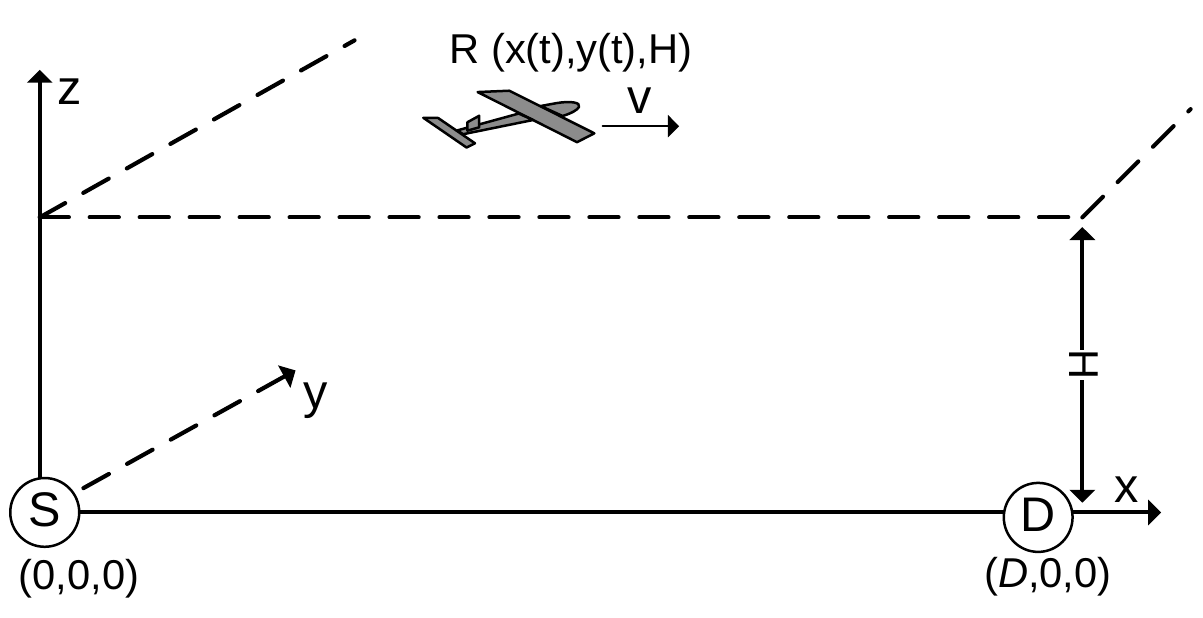}
\caption{Mobile relaying in 3D coordinate system.}\label{F:MobileRelay}
\end{figure}

\begin{itemize}
\item We present the basic model for mobile relaying in three dimensional (3D) Cartesian coordinate system, where a mobile relay with a given maximum speed as well as initial and final locations is employed to assist the communication from a source to a destination, as shown in Fig.~\ref{F:MobileRelay}. A throughput maximization problem is then formulated to optimize the  relay trajectory and the source/relay power allocations in a finite time horizon, subject to practical mobility, transmit power, and information-causality constraints.
\item Then, for fixed relay trajectory, we show that the optimal source/relay power allocations over time follow a ``staircase'' water-filling (WF) structure, with {\it non-increasing} and {\it non-decreasing} water levels at the source and relay, respectively. It is interesting to note that such a result is analogous to the optimal power allocation in {\it energy harvesting communications} \cite{532,636,635}, though they are owing to two different causality constraints, i.e., information-causality and energy-causality, respectively. Furthermore, for the particular relay trajectory such that  the source-relay and relay-destination channel gains are respectively non-increasing and non-decreasing over time, it is shown that the optimal source/relay power allocations reduce to the conventional WF solution with constant water levels, and either the source or relay should use up all its available transmit power.
\item Next, for a given source/relay power allocation, we propose an efficient algorithm to optimize the relay trajectory to further improve the throughput via applying successive convex optimization techniques. Specifically, the relay trajectory is successively updated  by finding the optimal trajectory incremental that maximizes a lower bound of the throughput. Based on the obtained results for separate power and trajectory optimizations, an iterative algorithm is then proposed to optimize both the power allocation and relay trajectory alternately.
\item Lastly, for the special case with free initial and final relay locations, we analytically derive the jointly optimal trajectory and power allocation solution for the throughput maximization problem.  In this case, it is shown that the relay with the optimal trajectory has only two states: either moves unidirectionally from the source to the destination with its maximum speed or stays stationary above the source or destination for a certain optimal duration.
\end{itemize}

It is worth pointing out that unlike existing {\it buffer-aided} static relaying techniques \cite{654}\cite{625}, which rely on random channel fading for opportunistic link selections to enhance performance, the proposed mobile relaying in this paper can pro-actively construct favorable channels via careful mobility control, and thus provides an additional degree of freedom for performance improvement.

The rest of this paper is organized as follows. Section~\ref{sec:systemModel} introduces the system model of mobile relaying, and presents the problem formulation for throughput maximization. In Section~\ref{sec:optSol}, the optimal source/relay power allocations are obtained for fixed relay trajectory. Section~\ref{sec:optTraj} optimizes the relay trajectory by assuming that the power allocations are fixed. In Section~\ref{sec:iterative}, an iterative algorithm is proposed to optimize both power allocation and relay trajectory by leveraging their individual optimized designs. In Section~\ref{sec:noRestr}, the jointly optimal relay trajectory and power allocation solution is analytically derived for the special case without pre-determined initial or final relay locations. In Section~\ref{sec:numerical}, numerical results are presented to compare the proposed mobile relaying design with  existing techniques. Finally, we conclude the paper and point out some future research directions in Section~\ref{sec:conclusion}.

\section{System Model and Problem Formulation}\label{sec:systemModel}
As shown in Fig.~\ref{F:MobileRelay}, we consider a wireless system with a source node $\Sn$ and a destination node $\D$ which are separated by $D$ meters. We assume that the direct link between $\Sn$ and $\D$ is negligible due to e.g., severe blockage. Thus, a relay $\R$ needs to be deployed to assist the communication from $\Sn$ to $\D$. Unlike the conventional static relaying with fixed relay location,  we assume that a relay of sufficiently high mobility is employed. In the following, we focus on the UAV-enabled mobile relaying, but the design principles are also applicable for other mobile relaying systems.

 Without loss of generality, we consider a three-dimensional (3D) Cartesian coordinate system with $\Sn$ and $\D$ located at $(0,0,0)$ and $(D,0,0)$, respectively, as shown in Fig.~\ref{F:MobileRelay}. We assume that a UAV  flying at a fixed altitude $H$ is employed as a mobile relay for a finite time horizon $T$. Thus, the time-varying coordinate of the relay node $\R$ can be expressed as $\big(x(t),y(t),H\big)$, $0 \leq t \leq T$, with $x(t)$ and $y(t)$ denoting the relay's time-varying x- and y-coordinates, respectively. Unless otherwise stated, we consider the scenario where the initial and final locations of the mobile relay are pre-determined, which are denoted as $(x_0, y_0, H)$ and $(x_F, y_F, H)$, respectively. This is because in practice, the initial and final relay locations depend on various factors such as the UAV's launching/landing locations as well as its pre- and post-mission flying paths, etc. In Section~\ref{sec:noRestr}, we also consider the case when the UAV is freely deployed to help relay information from $\Sn$ to $\D$, and as a result, there are no constraints on its initial and final locations. The minimum distance the relay needs to travel within the time horizon $T$ is $d_{\min}=\sqrt{(x_F-x_0)^2+(y_F-y_0)^2}$. Denote the maximum UAV speed as $\tilde{V}$, where $\tilde{V}\geq d_{\min}/T$ so that there exists at least one feasible trajectory from the relay's initial to final locations. We thus have $\sqrt{\dot{x}^2(t)+\dot{y}^2(t)}\leq \tilde{V}$, $0\leq t\leq T$, with $\dot{x}(t)$ and $\dot{y}(t)$ denoting the time-derivatives of $x(t)$ and $y(t)$, respectively.

 For ease of exposition, the time horizon $T$ is discretized into $N$ equally spaced time slots, i.e., $T=N\delta_t$, with $\delta_t$ denoting the elemental slot length, which is chosen to be sufficiently small so that the UAV's location can be assumed to be approximately constant within each slot. Thus, the UAV's trajectory $\big(x(t),y(t)\big)$ over $T$ can be approximated by the $N$-length sequences $\big\{x[n],y[n]\big\}_{n=1}^N$, where $\big(x[n],y[n]\big)$ denotes the UAV's x-y coordinate at slot $n$. As a result, the relay's mobility constraints, including both its initial and final location constraints as well as speed constraint,  can be expressed as
 \begin{align}
 \big(x[1]-x_0\big)^2+ \big(y[1]-y_0\big)^2\leq V^2, & \\
\big(x[n+1]-x[n]\big)^2+ \big(y[n+1]-y[n]\big)^2\leq V^2,&  \notag \\
n=1, \cdots, N-1,& \\
\big(x_F-x[N]\big)^2+ \big(y_F-y[N]\big)^2\leq V^2,&
 \end{align}
 where $V \triangleq \tilde{V}\delta_t$ denotes the maximum relay displacement for each time slot.

For simplicity, we assume that the relay $\R$ is equipped with a data buffer of sufficiently large size, and it operates in a frequency division duplexing (FDD) mode with equal bandwidth allocated for information reception from $\Sn$ and transmission to $\D$. Furthermore, we assume that the communication channels from $\Sn$ to $\R$ and that from $\R$ to $\D$ are dominated by line-of-sight (LoS) links, and the Doppler effect due to the relay's mobility is assumed to be perfectly compensated. Thus, at slot $n$, the channel power from $\Sn$ to $\R$ follows the free-space path loss model as 
 \begin{align}
 h_{\SR}[n]=\beta_{0} d_{\SR}^{-2}[n]=\frac{\beta_{0}}{H^2 + x^2[n]+y^2[n]},  \ n=1,\cdots, N,
 \end{align}
 where $\beta_{0}$ denotes the channel power at the reference distance $d_0=1$ meter, whose value depends on the carrier frequency, antenna gain, etc., and  $d_{\SR}[n]=\sqrt{H^2+ x^2[n]+y^2[n]}$ is the link distance between $\Sn$ and $\R$ at slot $n$. Let $p_s[n]$ denote the transmission power by $\Sn$ at slot $n$. The maximum transmission rate from $\Sn$ to $\R$ in bits/second/Hz (bps/Hz) for slot $n$ can be expressed as
 \begin{align}
R_{s}[n]&= \log_2\left(1+\frac{p_s[n] h_{\SR}[n]}{\sigma^2}\right),\notag \\
&= \log_2\left(1+\frac{p_s[n] \gamma_{0}}{H^2+ x^2[n]+y^2[n]}\right), n=1,\cdots, N,\label{eq:RSR}
 \end{align}
 where $\sigma^2$ denotes the noise power, and $\gamma_{0}\triangleq \beta_{0}/\sigma^2$ represents the reference signal-to-noise ratio (SNR). Similarly,  the channel from $\R$ to $\D$ at slot $n$ can be expressed as
$h_{\RD}[n]= \beta_{0}/(H^2+(D-x[n])^2+y^2[n])$, and the maximum transmission rate from $\R$ to $\D$ is
 \begin{align}
R_{r}[n]=\log_2\left(1+\frac{p_r[n]\gamma_{0}}{H^2+ (D-x[n])^2+y^2[n]}\right),& \notag \\
 n=1,\cdots, N,&\label{eq:RRD}
\end{align}
where $p_r[n]$ represents the transmission power by $\R$ at slot $n$. It follows from \eqref{eq:RSR} and \eqref{eq:RRD} that there in general exists a tradeoff in designing the relay trajectory $\{x[n]\}$ with given $\{y[n]\}$ between maximizing $\{R_s[n]\}$ versus $\{R_r[n]\}$ over the time slots.

Moreover, at each slot $n$, $\R$ can only forward the data that has already been received from $\Sn$. By assuming that the processing delay at $\R$ is one slot, we have the following {\it information-causality constraint}:
\begin{align}
R_{r}[1]=0, \ \sum_{i=2}^n R_{r}[i] \leq \sum_{i=1}^{n-1} R_{s}[i],\ n=2,\cdots, N.
\end{align}
It is not difficult to see that $\Sn$ should not transmit at the last slot $N$. We thus have $R_s[N]=R_r[1]=0$, and hence $p_s[N]=p_r[1]=0$ without loss of optimality.

For a given relay trajectory $\{x[n], y[n]\}_{n=1}^N$, define the time-dependent channel-to-noise power ratios  for the $\Sn$-$\R$ and $\R$-$\D$ links as
\begin{align}
&\gamma_{\SR}[n]\triangleq \frac{\gamma_{0}}{H^2+x^2[n]+ y^2[n]},\label{eq:gammaSR}\\
& \gamma_{\RD}[n]\triangleq \frac{\gamma_{0}}{H^2+(D-x[n])^2+ y^2[n]}, \forall n. \label{eq:gammaRD}
\end{align}
Our objective is to maximize the end-to-end throughput from $\Sn$ to $\D$ by optimizing both the source/relay power allocations $\{p_s[n]\}_{n=1}^{N-1}$ and $\{p_r[n]\}_{n=2}^N$ as well as the  relay trajectory $\{x[n], y[n]\}_{n=1}^N$. The problem can be formulated as follows.
\begin{align}
 & \mathrm{(P1):} \  \underset{\substack{\{x[n],y[n]\},\\ \{p_s[n],p_r[n]\}}}{\max}   \ \sum_{n=2}^N \log_2\Big(1+p_r[n]\gamma_{\RD}[n] \Big) \notag \\
  \text{s.t.}  &\  \sum_{i=2}^n \log_2\Big(1+p_r[i]\gamma_{\RD}[i]\Big) \leq \sum_{i=1}^{n-1} \log_2\Big(1+p_s[i]\gamma_{\SR}[i]\Big),\notag \\
   &\hspace{30ex} n=2,\cdots, N,\label{eq:InfoCausal2} \\
& \frac{1}{N}\sum_{n=1}^{N-1} p_s[n] \leq \bar P_s,\  \frac{1}{N} \sum_{n=2}^N p_r[n] \leq \bar P_r, \label{eq:PRConstr}  \\
& p_s[n] \geq 0, \ n=1,...,N-1, \\
&  p_r[n] \geq 0, \ n=2,...,N, \\
& \big(x[1]-x_0\big)^2+ \big(y[1]-y_0\big)^2\leq V^2,  \label{eq:initialConstr} \\
& \big(x[n+1]-x[n]\big)^2+ \big(y[n+1]-y[n]\big)^2\leq V^2,  \notag \\
& \hspace{30ex} n=1, \cdots, N-1, \label{eq:speedContr}\\
& \big(x_F-x[N]\big)^2+ \big(y_F-y[N]\big)^2\leq V^2, \label{eq:finalConstr}
\end{align}
where \eqref{eq:PRConstr} represents the average transmit power constraints over $T$, with $\bar P_s$ and $\bar P_r$ denoting the average power limits at $\Sn$ and $\R$, respectively. 


(P1) is a non-convex optimization problem, which thus cannot be directly solved with standard convex optimization techniques. In the following, we first consider two sub-problems of (P1), namely power optimization with fixed relay trajectory and trajectory optimization with fixed power allocation. Based on the solutions obtained, an iterative algorithm is then proposed for (P1) via alternately optimizing the  power and trajectory. Last, for the special case without pre-determined initial or final relay locations, i.e., in the absence of constraints \eqref{eq:initialConstr} and \eqref{eq:finalConstr}, we obtain the jointly optimal power allocation and relay trajectory solution to (P1).


\section{Power Optimization with Fixed Trajectory}\label{sec:optSol}
In this section, we consider the sub-problem of (P1) for optimizing the power allocations by assuming that the relay's trajectory $\{x[n],y[n]\}_{n=1}^N$ is fixed. Besides being a sub-problem of (P1), this may also correspond to the practical scenario when the relay's trajectory is pre-determined due to other tasks (e.g., surveillance) rather than being optimized for communication performance. In this case, it follows from \eqref{eq:gammaSR} and \eqref{eq:gammaRD} that the time-dependent channels $\{\gamma_{\SR}[n]\}$ and $\{\gamma_{\RD}[n]\}$ are given. However, the power allocation problem in the form of (P1) is still non-convex due to the non-convex information-causality constraints in \eqref{eq:InfoCausal2}. By introducing the slack variables $\{R_r[n]\}_{n=2}^N$, (P1) with given $\{\gamma_{\SR}[n]\}$ and $\{\gamma_{\RD}[n]\}$  can be reformulated as
 \begin{align}
 \hspace{-5ex} & \mathrm{(P1.1):}  \underset{\substack{\{p_s[n]\}_{n=1}^{N-1}, \\ \{p_r[n], R_r[n]\}_{n=2}^N}}{\max}   \ \sum_{n=2}^N R_r[n] \notag \\
  \text{s.t.}  &  \sum_{i=2}^n R_r[i] \leq \sum_{i=1}^{n-1} \log_2\Big(1+p_s[i]\gamma_{\SR}[i]\Big), n=2,\cdots,N\label{eq:InfoCausalConstr} \\
  &\ R_r[n] \leq \log_2\Big(1+p_r[n]\gamma_{\RD}[n]\Big),  n=2,\cdots,N \label{eq:RrConstr}\\
&\ \sum_{n=1}^{N-1} p_s[n] \leq E_s,\  \sum_{n=2}^N p_r[n] \leq E_r, \label{eq:PRConstr2} \\
&\ p_s[n] \geq 0, \ n=1,...,N-1, \label{eq:psConstr} \\
& \ p_r[n] \geq 0, \ n=2,...,N, \label{eq:prConstr}
\end{align}
where we have defined $E_s\triangleq NP_s$ and $E_r\triangleq NP_r$. Note that if at the optimal solution to (P1.1), there exists a slot $n'$ such that the constraint in \eqref{eq:RrConstr} is satisfied with strict inequality, we can always reduce the corresponding power $p_r[n']$ to make \eqref{eq:RrConstr} active, yet without decreasing the objective value of (P1.1). Thus, there always exists an optimal solution to (P1.1) such that all constraints in \eqref{eq:RrConstr} are satisfied with equality. As a result, for any fixed relay trajectory, (P1.1) is equivalent to (P1).  Note that (P1.1) is a convex optimization problem, which can be numerically solved by standard convex optimization techniques, such as the interior-point method \cite{202}. However, by applying the Lagrange dual method, the structural properties of the optimal solution to $\mathrm{(P1.1)}$ can be obtained, based on which new insights can be drawn.

\subsection{Optimal Solution to (P1.1)}
It can be verified that (P1.1) satisfies the Slater's condition, thus, strong duality holds and its optimal solution can be obtained via solving  the dual problem \cite{202}. Furthermore, the power and rate allocations for $\Sn$ and $\R$ in (P1.1) are only coupled  via the information-causality
 constraints in \eqref{eq:InfoCausalConstr}, which can be decoupled by studying its partial Lagrangian associated with this constraint.  Let $\lambda_n\geq 0$, $n=2,\cdots, N$, be the Lagrange dual variables corresponding to \eqref{eq:InfoCausalConstr}. The partial Lagrangian of (P1.1) can then be expressed as
\begin{align}
L &\left(\{p_s[n]\}, \{p_r[n], R_r[n], \lambda_n\} \right) \notag \\
=& \sum_{n=2}^N R_r[n] + \sum_{n=2}^N \lambda_n \left( \sum_{i=1}^{n-1} \log_2\left(1+p_s[i]\gamma_{\SR}[i]\right) -\sum_{i=2}^n R_r[i]\right)\notag \\
=& \sum_{n=2}^N \nu_n R_r[n] + \sum_{n=1}^{N-1}\beta_n \log_2\left(1+p_s[n]\gamma_{\SR}[n] \right),\label{eq:PartialL}
\end{align}
\begin{align}
\text{where} \hspace{10ex} &\beta_n \triangleq \sum_{i=n+1}^N \lambda_i, \ n=1,\cdots, N-1, \label{eq:betan}\\
&\nu_n\triangleq 1-\sum_{i=n}^N \lambda_i, \ n=2,\cdots, N. \label{eq:nun}
\end{align}

The Lagrange dual function of (P1.1) is then defined as
\begin{align}
g\left( \{\lambda_n\}\right) =
\begin{cases}
  \underset{\substack{\{p_s[n]\}_{n=1}^{N-1}, \\ \{p_r[n], R_r[n]\}_{n=2}^N}}{\max}  & \hspace{-2ex}  L \left(\{p_s[n]\}, \{p_r[n], R_r[n], \lambda_n\} \right) \notag \\
 \hspace{5ex} \text{s. t.}   & \hspace{-5ex} \eqref{eq:RrConstr}, \eqref{eq:PRConstr2}, \eqref{eq:psConstr}, \eqref{eq:prConstr}. \notag
\end{cases}
\end{align}
 The dual problem of (P1.1), denoted as (P1.1-D), is defined as $\min_{\lambda_n\geq 0, \forall n} g(\{\lambda_n\})$. Since (P1.1) can be solved equivalently by solving (P1.1-D), in the following, we first maximize the Lagrangian to obtain the dual function with fixed $\{\lambda_n\}$, and then find the optimal dual solutions $\{\lambda_n^\star\}$ to minimize the dual function. The optimal power and rate allocations at $\Sn$ and $\R$ are then obtained based on the dual optimal solution $\{\lambda_n^\star\}$.

Consider first the problem of maximizing the Lagrangian over $\{p_s[n]\}$ and $\{p_r[n], R_r[n]\}$ with fixed $\{\lambda_n\}$.
It follows from \eqref{eq:PartialL} that $g(\{\lambda_n\})$ can be decomposed as $g\left( \{\lambda_n\}\right)=g_s\left( \{\lambda_n\}\right)+g_r\left( \{\lambda_n\}\right)$, where
\begin{equation}
\begin{aligned}\label{eq:gs}
\hspace{-2ex} g_s\left( \{\lambda_n\}\right) =
\begin{cases}
\underset{\{p_s[n]\}}{\max} &  \sum_{n=1}^{N-1}\beta_n \log_2\left(1+p_s[n]\gamma_{\SR}[n] \right)   \\
 \text{s. t. }  & \sum_{n=1}^{N-1} p_s[n] \leq E_s, \\
  & \ p_s[n] \geq 0, \ n=1,...,N-1,
\end{cases}
\end{aligned}
\end{equation}
and
\begin{equation}
\begin{aligned}\label{eq:gr}
\hspace{-3ex} g_r\left( \{\lambda_n\}\right) =
\begin{cases}
\underset{\{p_r[n], R_r[n]\}}{\max} & \ \sum_{n=2}^{N}\nu_n R_r[n]  \\
 \text{s. t. }  &\hspace{-6ex} R_r[n] \leq \log_2\left(1+p_r[n]\gamma_{\RD}[n]\right), \forall n\\
 &\hspace{-6ex} \sum_{n=2}^{N} p_r[n] \leq E_r, \\
  &\hspace{-6ex} p_r[n] \geq 0, \ n=2,...,N.
  \end{cases}
\end{aligned}
\end{equation}
In other words, for any given dual variables $\{\lambda_n\}$, the optimal primal variables for Lagrangian maximization can be obtained by solving two parallel sub-problems \eqref{eq:gs} and \eqref{eq:gr} for $\Sn$ and $\R$, respectively. Note that both \eqref{eq:gs} and \eqref{eq:gr} are weighted sum-rate maximization problems each over $N-1$ parallel sub-channels, with the weights $\{\beta_n\}_{n=1}^{N-1}$ and $\{\nu_n\}_{n=2}^N$ determined by $\{\lambda_n\}_{n=2}^N$ given in \eqref{eq:betan} and \eqref{eq:nun}, respectively. Since $\lambda_n\geq 0$, $\forall n$, we have $\beta_n\geq 0$, $\forall n$, and $\{\beta_n\}_{n=1}^{N-1}$ and $\{\nu_n\}_{n=2}^N$ are {\it non-increasing} and {\it non-decreasing} over $n$, respectively. Furthermore, for problem \eqref{eq:gr} to have bounded optimal value, we must have $\nu_n\geq 0$, $\forall n$. To see this, suppose that there exists an $n'$ such that $\nu_{n'}<0$. Then problem \eqref{eq:gr} is unbounded when we let $R_r[n']=-t$, with $t\rightarrow \infty$. Since (P1.1) should have a bounded optimal value, it follows that the optimal primal and dual solutions of (P1.1) are obtained only when $\nu_n\geq 0$, $\forall n$, or equivalently $\sum_{n=2}^N \lambda_n \leq 1$ due to \eqref{eq:nun}.

By applying the standard Lagrange method and the Karush-Kuhn-Tucker (KKT) conditions, it is not difficult to show that the optimal solutions to \eqref{eq:gs} and \eqref{eq:gr} are respectively given by
\begin{align}
&p_s^\star[n]= \left[\eta \beta_n- \frac{1}{\gamma_{\SR}[n]} \right]^+, \ \forall n,\label{eq:ps} \\
& p_r^\star[n]= \left[ \xi \nu_n -\frac{1}{\gamma_{\RD}[n]}\right]^+, R_r^\star[n]= \left[\log_2 \left(\xi\nu_n \gamma_{\RD}[n] \right) \right]^+,  \forall n, \label{eq:pr}
\end{align}
where $\eta$ and $\xi$ are parameters ensuring $\sum_{n=1}^{N-1} p_s^\star[n]=E_s$ and $\sum_{n=2}^N p_r^\star[n]=E_r$, respectively, and $[a]^+\triangleq \max\{a,0\}$.

Next, we address how to solve the dual problem (P1.1-D) by minimizing the dual function $g(\{\lambda_n\})$ subject to $\lambda_n\geq 0$, $\forall n$, and the new constraint $\sum_{n=2}^N \lambda_n \leq 1$. This can be done by applying subgradient-based method, e.g., the ellipsoid method \cite{200}. It can be shown that the subgradient  of $g(\{\lambda_n\})$ at point $\{\lambda_n\}$ is given by $\mathbf s=[s_2, \cdots , s_N]^T$, with $s_n=\sum_{i=1}^{n-1}\log_2\left(1+p_s^\star[i]\gamma_{\SR}[i]\right)-\sum_{i=2}^n R_r^\star[i]$, $\forall n$, where $\{p_s^\star[n]\}$ and $\{R_r^\star[n]\}$ are the solutions in \eqref{eq:ps} and \eqref{eq:pr} for the given $\{\lambda_n\}$. The procedures for finding the optimal dual solutions $\{\lambda_n^\star\}$ using the ellipsoid method are summarized in Algorithm~\ref{Algo:primDual}.

With the dual optimal solution $\{\lambda_n^\star\}$ to (P1.1-D) obtained, the primal optimal solution to (P1.1), denoted as $\{p_s^{\opt}[n]\}$ and $\{p_r^{\opt}[n], R_r^{\opt}[n]\}$, can be obtained by separately considering the following four cases.

{\it Case 1: $\beta_1^\star>0$ and $\nu_N^\star>0$,} which is equivalent to $\sum_{n=2}^N \lambda_n^\star>0$ and $\lambda_N^\star<1$. In this case, both the weight vectors $\{\beta_n^\star\}$ in \eqref{eq:gs} and $\{\nu_n^\star\}$ in \eqref{eq:gr} have strictly positive components, and hence \eqref{eq:gs} and \eqref{eq:gr} are strict convex optimization problems and therefore have unique solutions. As a result, the solution given in \eqref{eq:ps} and \eqref{eq:pr} corresponding to the  dual optimal variable  $\{\lambda_n^\star\}$ must be the primal optimal solution to (P1.1). Note that in this case, $\Sn$ and $\R$ both use up their maximum transmission power. Furthermore, \eqref{eq:ps} and \eqref{eq:pr} show that the optimal power allocations  across the different slots are given by the ``staircase'' WF solution \cite{636}, with {\it non-increasing} and {\it non-decreasing} water levels at $\Sn$ and $\R$, respectively. 

{\it Case 2: $\beta_1^\star>0$ and $\nu_N^\star=0$,} or equivalently $\lambda^\star_N=1$ and $\lambda^\star_2=\cdots =\lambda^\star_{N-1}=0$. We then have $\beta_n^\star=1$, $\forall n$, and $\nu_n^\star=0$, $\forall n$. In this case, the weighted sum-rate maximization problem \eqref{eq:gs} reduces to sum-rate maximization problem, and its solution reduces to the classic WF power allocation with a constant water level \cite{209}, i.e., $p_s^\star[n]=\big[\eta-1/\gamma_{\SR}[n] \big]^+$, $\forall n$, with $\eta$ chosen such that $\sum_{n=1}^{N-1} p_s^\star[n]=E_s$. In this case, the unique Lagrangian maximizer $\{p_s^\star[n]\}$ must be the optimal power allocation for $\Sn$ corresponding to the primal optimal solution to (P1.1), i.e., $p_s^\opt[n]=p_s^\star[n]$, $\forall n$.
On the other hand, since $\nu_n^\star=0$, $\forall n$, problem \eqref{eq:gr} has non-unique solutions for Lagrangian maximization. The primal optimal solution can then be obtained by solving (P1.1) with the given optimal source power allocation $\{p_s^\opt[n]\}$. The resulting problem is a convex optimization problem of reduced complexity as compared to (P1.1).

Note that since $\lambda_N^\star=1$ for Case 2, the complementary slackness condition implies that $\sum_{n=2}^N R_r^\opt[n]=\sum_{n=1}^{N-1} R_s^\opt[n]$, i.e., the aggregated transmission rates at $\Sn$ and $\R$ are equal. Furthermore, as $\Sn$ (while not necessarily $\R$) must use up all its power to achieve such a rate balance, Case 2 corresponds to the scenario where the $\Sn$-$\R$ link is the bottleneck due to e.g., limited  power budget $E_s$ at $\Sn$ and/or poor channels $\{\gamma_{\SR}[n]\}$.

{\it Case 3: $\beta_1^\star=0$ and $\nu_N^\star>0$,} which corresponds to $\lambda_n^\star=0$, $\forall n$. Thus, we have $\beta_n^\star=0$, $\forall n$, and $\nu_n^\star=1$, $\forall n$. In this case, the optimal power allocation at $\R$ is given by the classic WF solution with a constant water level, i.e., $p_r^\opt[n]=\big[\xi-1/\gamma_{\RD}[n] \big]^+$, $\forall n$, with $\xi$ satisfying $\sum_{n=2}^{N} p_r^\star[n]=E_r$, and the resulting relay transmission rates are $R_r^\opt[n]= \left[\log_2 \left(\xi \gamma_{\RD}[n] \right) \right]^+$. On the other hand, as the source power allocation for the Lagrangian maximization \eqref{eq:ps} is not unique, we may obtain one as the primal optimal solution that minimizes the source transmission power while satisfying the information-causality constraint with the given relay transmission rates. 


{\it Case 4: $\beta_1^\star=0$ and $\nu_N^\star=0$}. This requires $\lambda_n^\star=0$, $\forall n$, on one hand, and also $\lambda_N^\star=1$ on the other hand. Thus, this case will not occur.

The complete algorithm for solving (P1.1) is summarized in Algorithm~\ref{Algo:primDual}.

\begin{algorithm}[H]
\caption{Optimal power allocation with fixed relay trajectory.}\label{Algo:primDual}
\begin{algorithmic}[1]
\STATE Initialize $\lambda_n\geq 0$, $\forall n$, and $\sum_{n=2}^N \lambda_n\leq 1$.
\REPEAT
\STATE Obtain $\{p_s^\star[n]\}$ and $\{p_r^\star[n], R_r^\star[n]\}$ using \eqref{eq:ps} and \eqref{eq:pr}.
\STATE Compute the subgradient of $g(\{\lambda_n\})$.
\STATE Update $\{\lambda_n\}$ using the ellipsoid method subject to $\lambda_n\geq 0$, $\forall n$ and $\sum_{n=2}^N \lambda_n\leq 1$.
\UNTIL{$\{\lambda_n\}$ converges to the prescribed accuracy}.
\STATE Output $\{p_s^\opt[n]\}$ and $\{p_r^\opt[n], R_r^\opt[n]\}$ according to the first three cases discussed above.
\end{algorithmic}
\end{algorithm}

\subsection{Optimal Power Allocation with Non-Increasing $\gamma_{\SR}[n]$ and Non-Decreasing $\gamma_{\RD}[n]$}
For the special case when the channels $\gamma_{\SR}[n]$ are $\gamma_{\RD}[n]$  are {\it non-increasing} and {\it non-decreasing} over $n$, respectively,  the optimal power allocation to (P1.1) can be obtained in closed-form. To this end, we first show the following result.

\begin{lemma}\label{lemma:zeroLambda}
If $\gamma_{\SR}[n]$ is non-increasing and $\gamma_{\RD}[n]$ is non-decreasing over $n$,  the  dual optimal solution $\{\lambda^\star_n\}$ to (P1.1) must satisfy $\lambda_n^\star=0$, $\forall n=2,\cdots, N-1$.
\end{lemma}
\begin{IEEEproof}
Please refer to Appendix~\ref{A:zeroLambda}.
\end{IEEEproof}

Note that Lemma~\ref{lemma:zeroLambda} only shows the vanishing of the dual variables associated with the information-causality constraints \eqref{eq:InfoCausalConstr} for slots up to $N-1$, whereas $\lambda_N^\star$ for the final slot could still be positive. In this case, it follows from \eqref{eq:betan} and \eqref{eq:nun} that $\beta_n^\star= \lambda^\star_N$, and $\nu_n^\star =1-\lambda^\star_N$, $\forall n$. As a result, the source and relay power allocations given in \eqref{eq:ps} and \eqref{eq:pr} with fixed dual optimal variables both reduce to the classic WF  solutions with constant water levels. With Lemma~\ref{lemma:zeroLambda}, the primal optimal solution to (P1.1) can be obtained in closed-form, as shown next.

For ease of presentation, we first define the following functions. For any $0\leq \tilde E_s\leq E_s$, define $\Rcwf_s(\tilde E_s)\triangleq \sum_{n=1}^{N-1}\left[\log_2\left( \eta \gamma_{\SR}[n]\right) \right]^+$ as the aggregated rate transmitted by $\Sn$ using the classic WF power allocation with total transmission power $\tilde E_s$, and $\pcwf_{s,n}(\tilde E_s)\triangleq \left[ \eta -1/\gamma_{\SR}[n]\right]^+$ as the corresponding power allocation for slot $n$, with $\eta$ satisfying $\sum_{n=1}^{N-1}\left[ \eta -1/\gamma_{\SR}[n]\right]^+=\tilde E_s$. Similarly, for $0\leq \tilde E_r \leq E_r$, define $\Rcwf_r(\tilde E_r)\triangleq \sum_{n=2}^N \left[\log_2\left(\xi \gamma_{\RD}[n]\right) \right]^+$, and $\pcwf_{r,n}(\tilde E_r)\triangleq \left[\xi-1/\gamma_{\RD}[n] \right]^+$, with $\xi$ satisfying $\sum_{n=2}^N \left[ \xi-1/\gamma_{\RD}[n]\right]^+=\tilde E_r$. We then have the following result.

\begin{theorem}\label{theo:th1}
If $\gamma_{\SR}[n]$ is non-increasing and $\gamma_{\RD}[n]$ is non-decreasing over $n$, an optimal power allocation to (P1.1) is
$p_s^\opt[n]=\pcwf_{s,n} ( \tilde E_s^\opt), \ p_r^\opt[n]=\pcwf_{r,n}( \tilde E_r^\opt), \ \forall n$,
\begin{align}
\text{where }
\big(\tilde E_s^\opt, \tilde E_r^\opt\big) =
\begin{cases}
\big(E_s, \hat E_r\big) \ & \text{ if }  \Rcwf_s(E_s) \leq \Rcwf_r(E_r) \\
 \big(\hat E_s,  E_r\big), \ & \text { otherwise},\notag
\end{cases}
\end{align}
with $\hat E_s$ and $\hat E_r$ denoting the unique solution to the equation $\Rcwf_s(\tilde E_s)=\Rcwf_r(E_r)$ and $\Rcwf_r(\tilde E_r)=\Rcwf_s(E_s)$, respectively. Furthermore, the corresponding optimal value of (P1.1) is
\begin{align}
R^\opt = \min\{\Rcwf_s(E_s), \Rcwf_r(E_r)\}.\label{eq:Ropt}
\end{align}
\end{theorem}
\begin{IEEEproof}
Please refer to Appendix~\ref{A:th1}.
\end{IEEEproof}

Theorem~\ref{theo:th1} states that if the relay moves unidirectionally from $\Sn$ to $\D$ so that $\gamma_{\SR}[n]$  and $\gamma_{\RD}[n]$ are non-increasing and non-decreasing over $n$, respectively,  the optimal power allocations at both $\Sn$ and $\R$ reduce to the classic WF solution with optimized total transmit power $\tilde E_s^\opt$ and $\tilde E_r^\opt$, respectively. Specifically, by ignoring the information-causality constraints \eqref{eq:InfoCausalConstr}, the transmitter corresponding to the ``bottleneck'' link  which has smaller aggregate rate $\Rcwf_s(E_s)$ or $\Rcwf_r(E_r)$ should use up all its available power, whereas the other transmitter may reduce its power so as to balance the rates over the two links. Under such transmission strategies, the information-causality constraints are automatically guaranteed, which is intuitively understood since the $\Sn$-$\R$ link always has better channels, and hence higher power and rate, in earlier slots, whereas the reverse is true for the $\R$-$\D$ link.


\section{Trajectory Optimization with Fixed Power}\label{sec:optTraj}
In this section, we consider another sub-problem of (P1) for optimizing the relay's trajectory $\{x[n],y[n]\}_{n=1}^N$ with fixed source and relay power allocations $\{p_s[n]\}_{n=1}^{N-1}$ and $\{p_r[n]\}_{n=2}^N$. Notice that this sub-problem is particularly relevant when the relay and source can only transmit  with constant power due to practical hardware limitations.  The problem can be written as
\begin{align}
 & \mathrm{(P1.2):}  \underset{\substack{\{x[n],y[n]\}_{n=1}^N\\ \{R_r[n]\}_{n=2}^N}}{\max}  \sum_{n=2}^N R_r[n] \notag \\
  \text{s.t.}  &  \sum_{i=2}^n R_r[i] \leq \sum_{i=1}^{n-1} \log_2\left(1+\frac{\gamma_s[i]}{H^2+ x^2[i]+y^2[i]}\right), \notag \\
  & \hspace{30ex} n=2,\cdots, N, \label{eq:InfoCausalConstr2} \\
  & R_r[n] \leq \log_2 \left(1+\frac{\gamma_r[n]}{H^2+ (D-x[n])^2 + y^2[n]}\right), \notag \\
  & \hspace{30ex} n=2,\cdots, N, \label{eq:RrConstr2}\\
& \big(x[1]-x_0\big)^2+ \big(y[1]-y_0\big)^2\leq V^2,   \\
& \big(x[n+1]-x[n]\big)^2+ \big(y[n+1]-y[n]\big)^2\leq V^2,  \notag \\
& \hspace{30ex} n=1, \cdots, N-1,  \\
& \big(x_F-x[N]\big)^2+ \big(y_F-y[N]\big)^2\leq V^2,
\end{align}
where $R_r[n]$ is the slack variable denoting the relay's transmission rate at slot $n$, $\gamma_s[n]\triangleq p_s[n]/\sigma^2$ and $\gamma_r[n]\triangleq p_r[n]/\sigma^2$, $\forall n$.

(P1.2) is a non-convex optimization problem due to the non-convex constraints \eqref{eq:InfoCausalConstr2} and \eqref{eq:RrConstr2}. Therefore, it is quite challenging to find its optimal solution efficiently. In the following, we obtain an efficient approximate solution to (P1.2) based on the successive convex optimization technique. The main idea is to successively maximize a lower bound of (P1.2) via optimizing the incremental of the relay's trajectory at each iteration. Specifically, let $\{x_l[n], y_l[n]\}_{n=1}^N$ be the resulting relay trajectory after the $l$th iteration, and $R_{s,l}[n]\triangleq \log_2\left(1+\frac{\gamma_s[n]}{H^2+ x_l^2[n]+y_l^2[n]}\right)$ and $R_{r,l}[n]\triangleq \log_2 \left(1+\frac{\gamma_r[n]}{H^2+ (D-x_l[n])^2 + y_l^2[n]}\right)$ be the corresponding channel capacity for the $\Sn$-$\R$ and $\R$-$\D$ links, respectively. Further denote $\{\delta_l[n], \xi_l[n]\}_{n=1}^N$ as the trajectory incremental from the $l$th to the $(l+1)$th iteration, i.e., $x_{l+1}[n]=x_{l}[n]+\delta_l[n]$, $y_{l+1}[n]=y_{l}[n]+\xi_{l}[n]$, $\forall n$.
  We then have the following result.

  \begin{lemma}\label{lemma:approx}
  For any trajectory incremental $\{\delta_{l}[n]\}$ and $\{\xi_l[n]\}$, the following inequalities hold
 \begin{align}
 R_{s, l+1}[n]  \geq & R_{s,l+1}^{\LB}[n]\triangleq R_{s,l}[n] - a_{s,l}[n] \big(\delta_l^2[n] + \xi_{l}^2[n] \big) \notag \\
  & - b_{s,l}[n] \delta_l[n] - c_{s,l}[n] \xi_l[n], \label{eq:RsNext}\\
 R_{r, l+1}[n]  \geq & R_{r,l+1}^{\LB}[n] \triangleq R_{r,l}[n] - a_{r,l}[n] \big(\delta_l^2[n] + \xi_{l}^2[n] \big) \notag \\
 & - b_{r,l}[n] \delta_l[n] - c_{r,l}[n] \xi_l[n], \forall n, \label{eq:RrNext}
 \end{align}
 where $a_{s,l}[n]$, $a_{r,l}[n]\geq 0$, $b_{s,l}[n]$, $c_{s,l}[n]$,  $b_{r,l}[n]$, and $c_{r,l}[n]$ are coefficients given by \eqref{eq:CoeffS} and \eqref{eq:CoeffR} of Appendix~\ref{A:approx}.
 \end{lemma}
\begin{IEEEproof}
Please refer to Appendix~\ref{A:approx}.
\end{IEEEproof}

 Lemma~\ref{lemma:approx} shows that for any existing relay trajectory $\{x_l[n], y_l[n]\}$ and an additional trajectory incremental $\{\delta_l[n], \xi_l[n]\}$, the resulting new channel capacity $R_{s,l+1}[n]$ and $R_{r,l+1}[n]$ are lower-bounded by $ R_{s,l+1}^{\LB}[n]$ and $ R_{r,l+1}^{\LB}[n]$, respectively, which are concave quadratic functions of $\delta_l[n]$ and $\xi_{l}[n]$ since $a_{s,l}[n], a_{r,l}[n]\geq 0$. It then follows that the optimal value of (P1.2), denoted as $R^\opt$, is lower-bounded by that of the following problem for any given trajectory $\{x_l[n], y_l[n]\}$, 
 \begin{align}
 \hspace{-5ex} & \mathrm{(P1.3):}  \underset{\substack{\{\delta_l[n],\xi_l[n]\}_{n=1}^N\\ \{R_r[n]\}_{n=2}^N}}{\max}  \sum_{n=2}^N R_r[n] \notag \\
  \text{s.t.}  &  \sum_{i=2}^n R_r[i] \leq \sum_{i=1}^{n-1} 
    R_{s,l+1}^\LB[i], \ n=2,\cdots, N, \label{eq:InfoCausalConstr4} \\
  & R_r[n] \leq  
  R_{r,l+1}^\LB[n], \ n=2,\ \cdots, N, \label{eq:RrConstr4}  \\
& \big(x_l[1]+\delta_l[1]-x_0\big)^2+ \big(y_l[1]+\xi_l[1]-y_0\big)^2\leq V^2,   \\
& \big(x_l[n+1]+\delta_l[n+1]-x_l[n]-\delta_l[n]\big)^2+ \notag \\
& \hspace{2ex} \big(y_l[n+1]+\xi_l[n+1]-y_l[n]-\xi_l[n]\big)^2\leq V^2, \forall n,\\
& \big(x_F-x_l[N]-\delta_l[N]\big)^2+ \big(y_F-y_l[N]-\xi_l[N]\big)^2\leq V^2.
\end{align}

(P1.3) is a convex quadratic programming problem, which thus can be efficiently solved with the standard convex optimization technique or existing software tools such as CVX \cite{227}. As a result, (P1.2) can then be approximately solved by successively updating the trajectory based on the optimal solution to (P1.3), which is summarized in Algorithm~\ref{Algo:SCA}.

\begin{algorithm}[H]
\caption{Successive trajectory optimization with fixed power allocation.}\label{Algo:SCA}
\begin{algorithmic}[1]
\STATE Initialize the relay's trajectory as $\{x_0[n], y_0[n]\}_{n=1}^N$, and let $l=0$.
\REPEAT
\STATE Find the optimal solution $\{\delta_l^\star[n], \xi_l^\star[n]\}_{n=1}^N$ to (P1.3).
\STATE Update the trajectory $x_{l+1}[n]=x_l[n]+\delta_l^\star[n]$ and $y_{l+1}[n]=y_l[n]+\xi_l^\star[n]$, $\forall n=1,\cdots, N$.
\STATE Update $l=l+1$.
\UNTIL{convergence or a maximum number of iterations has been reached.}
\end{algorithmic}
\end{algorithm}
It can be shown that with Algorithm~\ref{Algo:SCA}, the resulting optimal values of (P1.3) are non-decreasing over the iteration $l$, which are further upper-bounded by the optimal value of (P1.2). Thus, Algorithm~\ref{Algo:SCA} is guaranteed to converge. 

\section{Iterative Power and Trajectory Optimization}\label{sec:iterative}
In this section, we propose an iterative algorithm for the joint power and trajectory optimization problem (P1) based on the solutions to its two  sub-problems  obtained in the preceding two sections. The main idea is to alternately optimize the power allocation and the relay's trajectory by assuming that the other design variable is fixed.  The algorithm is summarized in Algorithm~\ref{Algo:iterative}.

\begin{algorithm}[H]
\caption{Iterative power and trajectory optimization.}\label{Algo:iterative}
\begin{algorithmic}[1]
\STATE Initialize the relay's trajectory.
\REPEAT
\STATE Fix the relay's trajectory, find the optimal power allocations using Algorithm~\ref{Algo:primDual}.
\STATE Fix the power allocation, update the relay's trajectory using Algorithm~\ref{Algo:SCA}.
\UNTIL{convergence or a maximum number of iterations has been reached.}
\end{algorithmic}
\end{algorithm}

Note that as each iteration of Algorithm~\ref{Algo:iterative} only requires solving convex optimization problems, the overall complexity of Algorithm~\ref{Algo:iterative} is polynomial in the worst scenario. However, since the sub-problem (P1.2) for trajectory optimization cannot be guaranteed to be optimally solved by Algorithm~\ref{Algo:SCA}, no optimality can be theoretically declared for Algorithm~\ref{Algo:iterative}. 
However,  for the special case without pre-determined initial or final relay locations, where the jointly optimal solution to (P1) can be analytically obtained as shown in the next section, the numerical results in Section~\ref{sec:numerical} show that Algorithm~\ref{Algo:iterative} yields near optimal performance.

\section{Optimal Solution with Free Initial/Final Relay Location}\label{sec:noRestr}
In this section, we derive the jointly optimal solution to (P1) for the particular case when there is no pre-specified initial or final relay location. In practice, this could correspond to the scenario where the UAV is dedicated to assist communication and thus can be launched/landed in any optimized location via e.g., ground transportation before mission starts and after mission is completed. In this case, (P1) is solved by removing the constraints \eqref{eq:initialConstr} and \eqref{eq:finalConstr}. The resulting problem is denoted as (P1'). We first present the following result.

\begin{lemma}\label{lemma:SimpleLemma}
Without loss of optimality to (P1'), we have $0\leq x[n] \leq D$ and $y[n]=0$, $\forall n$.
\end{lemma}
\begin{IEEEproof}
First, it is obvious that $\{y[n]\}$ should be all equal to zeros, since otherwise, both channels in \eqref{eq:gammaSR} and \eqref{eq:gammaRD} can be improved and the feasible region for $\{x[n]\}$ in \eqref{eq:speedContr} can be enlarged by setting $\{y[n]\}$ equal to zeros. Also,  it follows from \eqref{eq:gammaSR} and \eqref{eq:gammaRD} that $0\leq x[n] \leq D$, since otherwise, we can always find an alternative relay location within the interval $[0, D]$ that results in higher $\gamma_{\SR}[n]$ and/or $\gamma_{\RD}[n]$.
\end{IEEEproof}

To obtain the optimal solution to (P1'), we first show that the optimal relay trajectory $\{x[n]\}$ is {\it non-decreasing} over $n$, i.e., the relay should move unidirectionally towards $\D$. As a result, it then follows from  Lemma~\ref{lemma:SimpleLemma} that the channels $\gamma_{\SR}[n]$ and $\gamma_{\RD}[n]$ in \eqref{eq:gammaSR} and \eqref{eq:gammaRD} are non-increasing and non-decreasing, respectively. Therefore, the optimal power allocations can be obtained in closed-form given by Theorem~\ref{theo:th1}. With slight abuse of notations, we first denote $\Rcwf_s(E_s)$ and $\Rcwf_r(E_r)$ in \eqref{eq:Ropt} as $\Rcwf_s(\{x[n]\})$ and $\Rcwf_r(\{x[n]\})$, i.e., as the functions of the relay trajectory $\{x[n]\}$ explicitly.

\begin{theorem}\label{thm:nonIncreasingxn}
Without loss of optimality to (P1'), the relay trajectory $\{x[n]\}$ is non-decreasing over $n$.
\end{theorem}
\begin{IEEEproof}
Please refer to Appendix~\ref{A:nonIncreasingxn}.
\end{IEEEproof}

It  then follows from Theorem~\ref{theo:th1} and Theorem~\ref{thm:nonIncreasingxn} that problem (P1') for joint power and trajectory optimization reduces to determining the optimal relay trajectory $\{x[n]\}$ by solving
\begin{align}
\mathrm{(P2):}\ \underset{\{x[n]\}}{\max} & \ \min \Big \{ \Rcwf_r\left(\{x[n]\} \right), \Rcwf_s\left(\{x[n]\} \right)  \Big \} \notag \\
\text{s.t.} & \ 0 \leq x[n+1]-x[n] \leq V, \ \forall n \label{eq:mobileConstr1} \\
&\ 0\leq x[n] \leq D, \ \forall n. \label{eq:mobileConstr}
\end{align}
where \eqref{eq:mobileConstr1} follows from the speed constraint \eqref{eq:speedContr}  by applying both Lemma~\ref{lemma:SimpleLemma} and Theorem~\ref{thm:nonIncreasingxn}.

\begin{theorem}\label{theo:th3}
Without loss of optimality to (P2), $\{x[n]\}$ satisfies
\begin{align}
v[n]=
\begin{cases}
V, & \text{ if } 0 < x[n] <D, \\
0, & \text{ if } x[n]=D,\\
V \text{ or } 0, & \text{if } x[n]=0,
\end{cases}
\end{align}
where $v[n]\triangleq x[n+1]-x[n]$ is the velocity at slot $n$.
\end{theorem}
\begin{IEEEproof}
Please refer to Appendix~\ref{A:th3}.
\end{IEEEproof}

Theorem~\ref{theo:th3} shows that  a binary decision on the velocity with $v[n]$ equal to either $0$ or $V$ is optimal to (P2). Furthermore, the relay stays stationary, i.e., $v[n]=0$, only if $x[n]=0$ or $x[n]=D$, when it enjoys the best channel either from the source or to the destination. As a result, (P2) can be optimally solved by considering the following four scenarios. 

\begin{figure}
\centering
\includegraphics[scale=0.45]{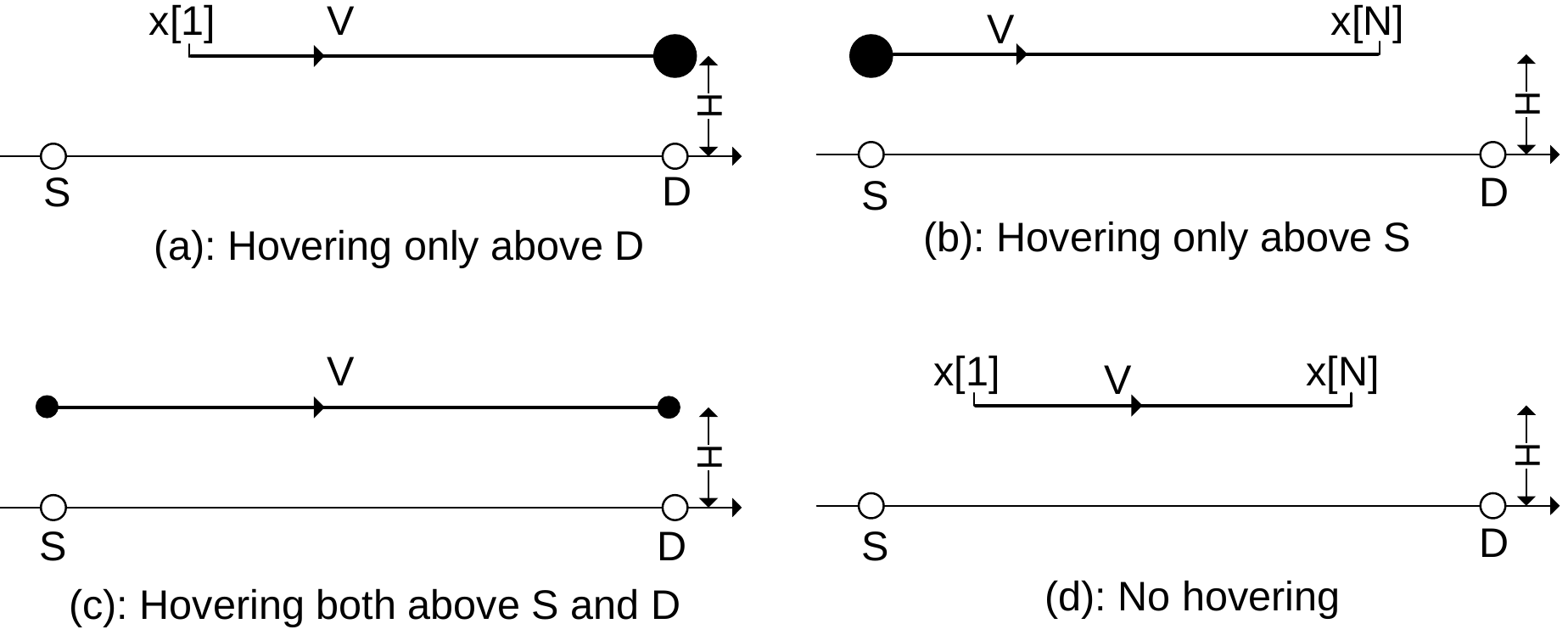}
\caption{Four scenarios of optimal relay trajectory for (P1').}\label{F:TrajectoryFourCases}
\end{figure}

\subsubsection{Scenario (a), Hovering only above $\D$}
As illustrated in Fig.~\ref{F:TrajectoryFourCases}(a), in this scenario, $\R$ moves from a starting position $x[1]\geq 0$ towards $\D$ with the maximum speed $V$, and remains stationary after it reaches $\D$. Thus, the relay trajectory can be parameterized by $x[1]$ as $x[n]=\big[x[1]+(n-1)V\big]_0^D$, $\forall n$, where $[\cdot]_a^b$ represents projection into the interval $[a, b]$.  As a result, (P2) reduces to determining the optimal starting position $x[1]$. Since $\Rcwf_s(\cdot)$ and $\Rcwf_r(\cdot)$ are respectively {\it non-increasing} and {\it non-decreasing} functions over $x[1]$, the optimal $x[1]$ to (P2) can be efficiently obtained via bisection search over the interval $[0, D]$.

\subsubsection{Scenario (b), Hovering only above $\Sn$}
As illustrated in Fig.~\ref{F:TrajectoryFourCases}(b), in this scenario, starting from $\Sn$, $\R$ hovers above $\Sn$ for some duration (if $N$ is sufficiently large), and moves towards $D$ with the maximum speed. In this case, the trajectory can be parameterized by the final position $x[N]$ as $x[n]=\big[x[N]-(N-n)V\big]_0^D$, $\forall n$. Similar to scenario (a), the optimal $x[N]$ to (P2) can be efficiently obtained via bisection method.

\subsubsection{Scenario (c), Hovering both above $\Sn$ and $\D$}
As illustrated in Fig.~\ref{F:TrajectoryFourCases}(c), in this scenario, $\R$ moves from $\Sn$ to $\D$ with the maximum speed, and remains stationary for some durations when it is both above $\Sn$ and $\D$. Thus, the trajectory can be expressed as
 \begin{align}\label{eq:Case3}
 x[n]= \begin{cases}0, & \ 1 \leq n \leq N_1 \\
 V(n-N_1), & \ N_1 < n \leq N_1+\frac{D}{V} \\
 D, & \ N_1+\frac{D}{V}< n \leq N,
 \end{cases}
 \end{align}
 where $N_1$ is the number of slots for $\R$ hovering above $\Sn$.  Note that this case is possible only if the speed $V$ is sufficiently large such that $NV>D$. With \eqref{eq:Case3}, (P2) reduces to determining the optimal $N_1$. As $\Rcwf_s(\cdot)$ and $\Rcwf_r(\cdot)$ are respectively non-decreasing and non-increasing functions over $N_1$, the optimal $N_1$ to (P2) can be efficiently obtained by bisection method.

 \subsubsection{Scenario (d), Hovering neither above $\Sn$ nor $\D$}
 It can be shown that this scenario is a special case of  Scenario (a) or (b). Thus, no separate optimization is needed.

 The optimal solution to (P2), and hence the jointly optimal solution to (P1'), is then obtained by comparing the optimal values corresponding to the first three scenarios discussed above.


%

 \begin{figure*}
\centering
\includegraphics[scale=0.6]{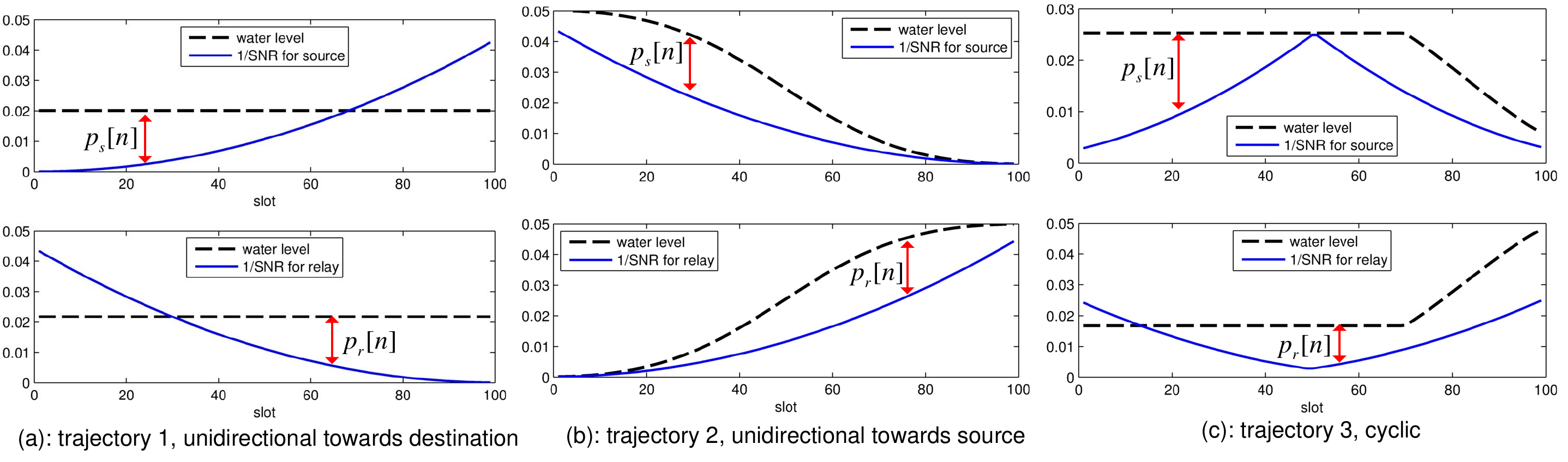}
\caption{Optimal power allocation for three different UAV trajectories.}\label{F:PowerAllocations}
\end{figure*}

\section{Numerical Results}\label{sec:numerical}
In this section, numerical results are provided to validate our proposed mobile relaying design.  We consider a system with the source $\Sn$ and the destination $\D$ separated by $D=2000$m. The communication bandwidth per link is $20$MHz with the carrier frequency at $5$GHz, and the noise power spectrum density is  $-169$dBm/Hz. Thus, the reference SNR at the distance $d_0=1$m can be obtained as $\gamma_{0}=80$dB. For the mobile relaying system, the maximum UAV speed is assumed to be $\tilde{V}=50$m/s, and its flying altitude is fixed to $H=100$m. For the benchmark static relaying system, the relay is assumed to be fixed at the location $(D/2, 0, H)$. Unless otherwise specified, the maximum average transmit power at both $\Sn$ and $\R$ is assumed to be $\bar P_s=\bar P_r=10$ dBm.

\subsection{Power Optimization with Fixed Trajectory}
First, we consider the mobile relaying system with fixed relay trajectory, whereas the power allocations at the source and relay are optimized as in Section~\ref{sec:optSol}. We consider three specific UAV/relay trajectories: (a) {\it unidirectional towards $\D$}, for which the UAV moves unidirectionally from $\Sn$ to $\D$ with the maximum speed; (b) {\it unidirectional towards $\Sn$}, where the UAV moves in the reverse direction from $\D$ to $\Sn$ with the maximum speed; (c) {\it cyclic between $D/4$ and $3D/4$}.  Fig.~\ref{F:PowerAllocations} illustrates the optimal power allocations at $\Sn$ and $\R$ over different slots for the three trajectories.  It is observed from Fig.~\ref{F:PowerAllocations}(a) that for unidirectional movement to $\D$, the power allocations at both $\Sn$ and $\R$ follow the classic WF with a certain constant water level, which is in accordance with Theorem~\ref{theo:th1}; whereas for Fig.~\ref{F:PowerAllocations}(b) with the reverse movement, the water levels at $\Sn$ and $\R$ keep decreasing and increasing, respectively, which implies that the information-causality constraint is always active, i.e., the received data at $\R$ is immediately forwarded at the subsequent slot. For the cyclic movement shown in Fig.~\ref{F:PowerAllocations}(c), the water levels at both $\Sn$ and $\R$ are initially constant, and then decrease and increase respectively after certain time.

\begin{figure}
\centering
\includegraphics[scale=0.6]{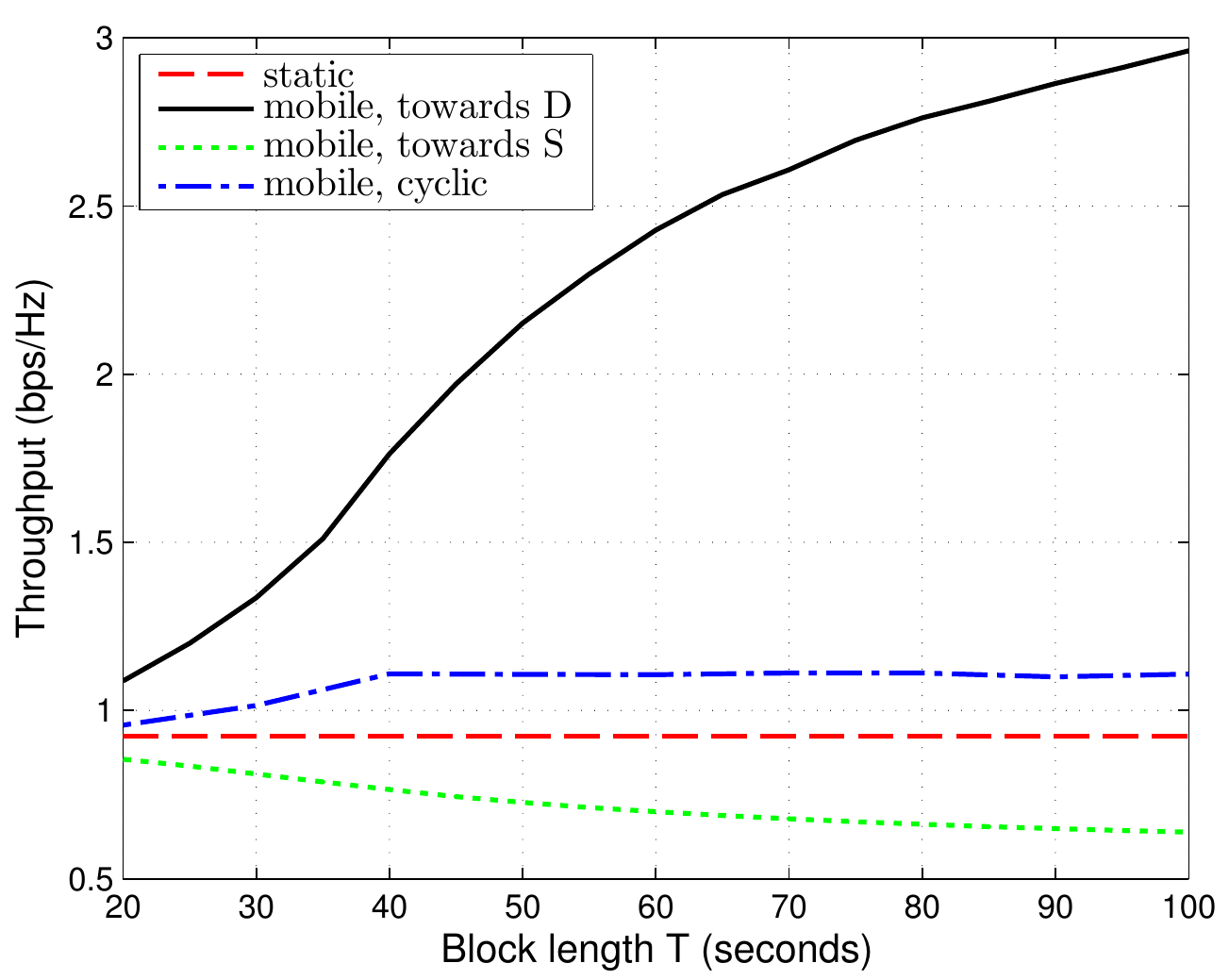}
\caption{Throughput of static versus mobile relaying with different UAV trajectories.}\label{F:ThroughputVSTFixedTrajectory}
\end{figure}

In Fig.~\ref{F:ThroughputVSTFixedTrajectory}, the throughput in bps/Hz versus the duration $T$ is plotted for the static versus mobile relaying with the three aforementioned relay trajectories. Note that when $T$ is sufficiently large, the UAV for the two unidirectional schemes could stay stationary above $\Sn$ (and above $\D$) for certain period before it moves towards $\D$ (after it arrives above $\D$). It is observed that with the UAV moving unidirectionally towards $\D$, the mobile relaying scheme significantly outperforms the conventional static relaying, thanks to the reduced link distances for both information reception and forwarding by relay movement from $\Sn$ to $\D$. In contrast, for unidirectional relay movement from $\D$ to $\Sn$, the performance is even worse than the conventional static relaying. This is expected since with this specific relay movement, both $\Sn$ and $\R$ are forced to allocate more power on weak channels due to the information-causality constraint, as can be seen from Fig.~\ref{F:PowerAllocations}(b). Such results imply the necessity of joint UAV trajectory and power allocations in order to realize the full benefit of mobile relaying technique.

 \begin{figure}
\centering
\includegraphics[scale=0.6]{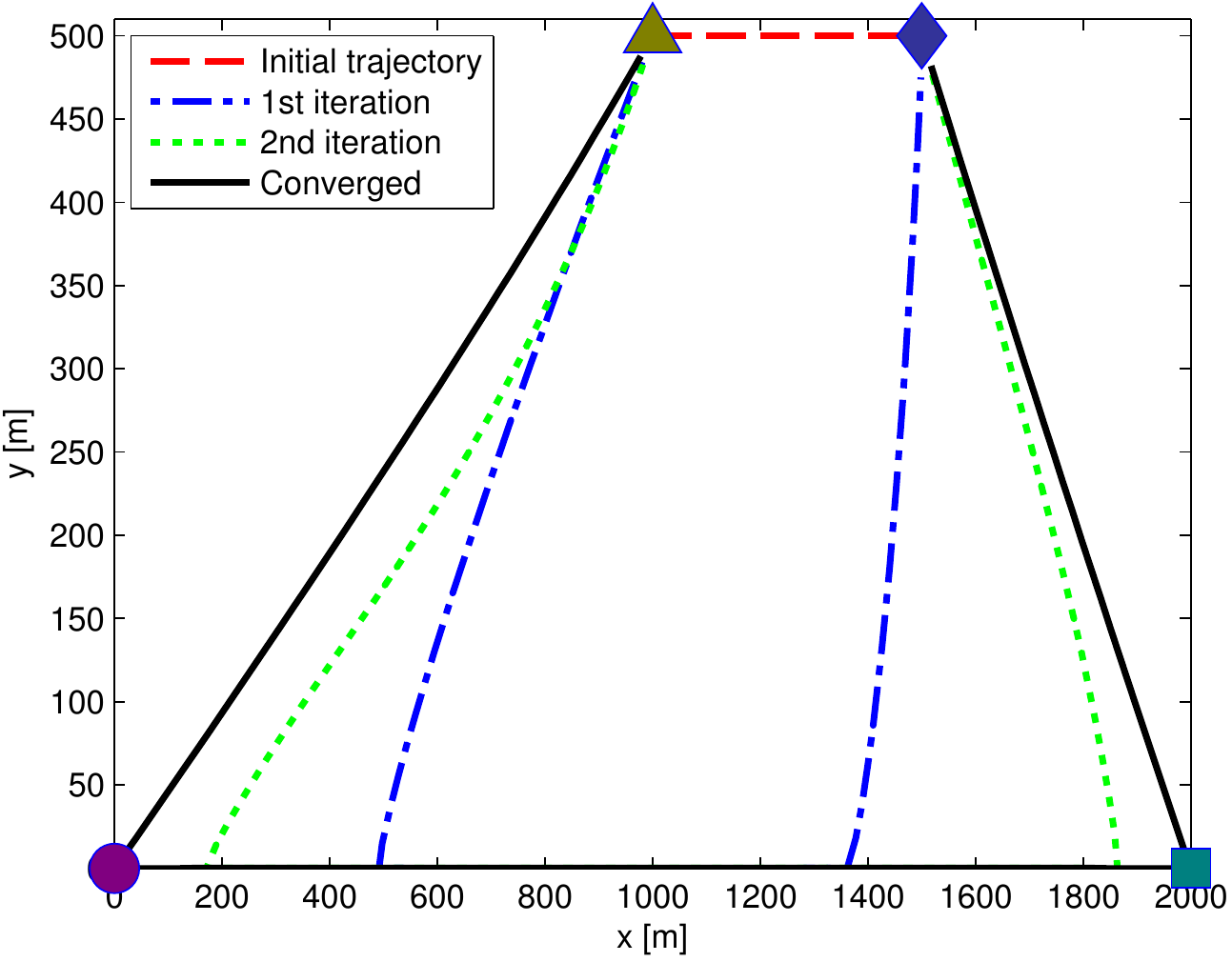}
\caption{UAV trajectory evolution by Algorithm~\ref{Algo:SCA}. The circle, square, triangle,  and diamond represent the source, destination, and initial and final relay locations, respectively.}\label{F:TrajectoryUpdatesV50T40FixPowerEqualPower}
\end{figure}

\subsection{Trajectory Optimization with Fixed Power Allocation}
Next, we consider the mobile relaying system where the power allocations at the source and relay over different time slots are fixed, whereas the relay's trajectory is optimized as in Section~\ref{sec:optTraj}. We assume that the relay's initial and final x-y coordinates are pre-determined and given by $(x_0,y_0)=(1000,500)$ and $(x_F, y_F)=(1500, 500)$, respectively, as shown in Fig.~\ref{F:TrajectoryUpdatesV50T40FixPowerEqualPower}. Therefore, the minimum distance that the relay needs to travel within the time horizon $T$ is $d_{\min}=500$m. We assume that equal power allocation across different time slots is applied at both the source and relay, and  Algorithm~\ref{Algo:SCA} is applied to successively optimize the relay trajectory, where the initial trajectory is set to be the direct path from $(x_0,y_0)$ to $(x_F, y_F)$ with uniform traveling speed.

\begin{figure}
\centering
\includegraphics[scale=0.52]{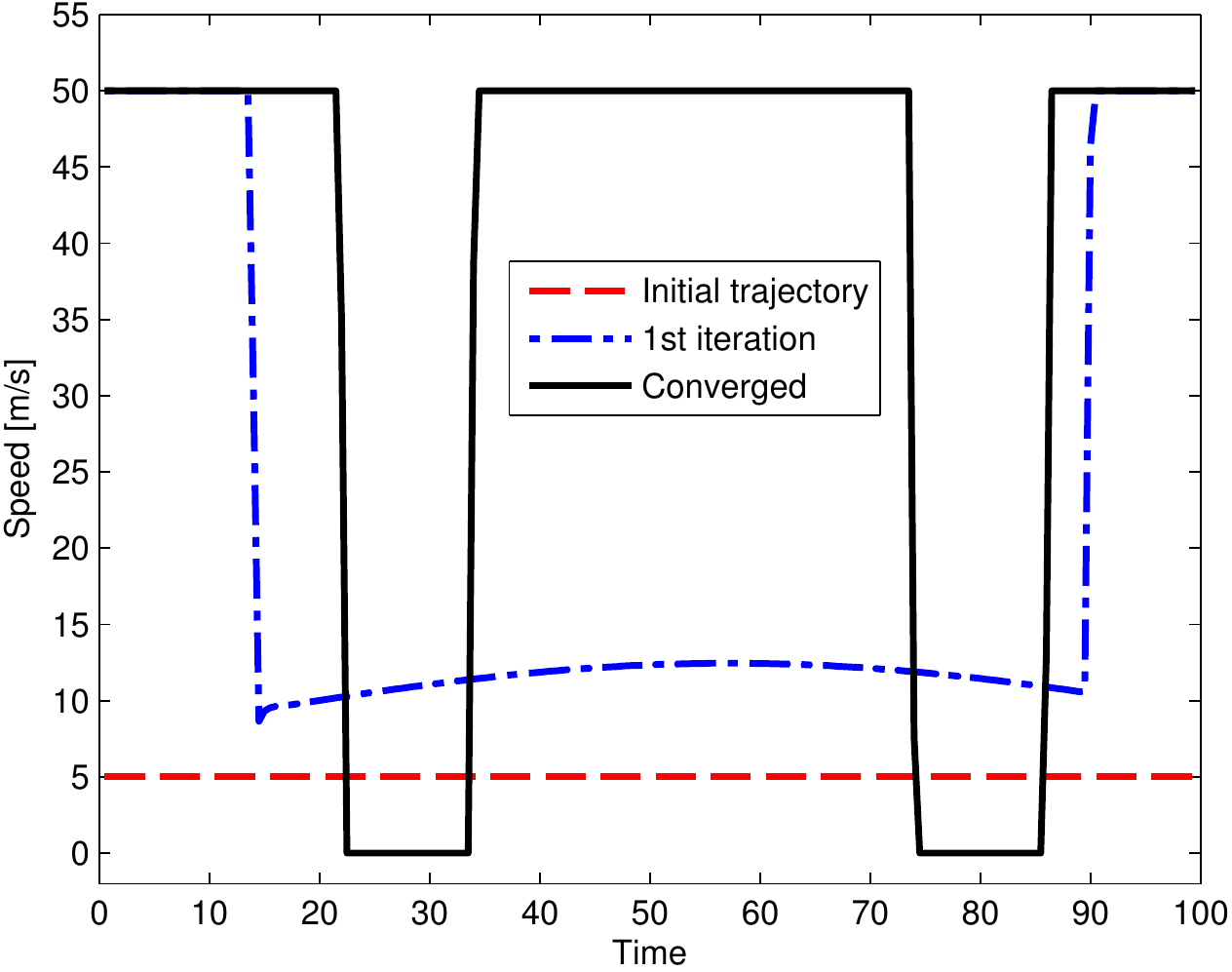}
\caption{The speed of the mobile relay over time for three different trajectories from Fig.~\ref{F:TrajectoryUpdatesV50T40FixPowerEqualPower}.}\label{F:speed}
\end{figure}

For $T=100$s, Fig.~\ref{F:TrajectoryUpdatesV50T40FixPowerEqualPower} shows the projected relay trajectories onto the horizontal plane obtained with different iterations of Algorithm~\ref{Algo:SCA}. It is observed that instead of following the direct path, the optimized trajectory first moves towards $\Sn$ and then to $\D$ before heading towards its final location. This is expected since the fact that $\tilde V T > d_{\min}$ offers the degree of freedom for dynamically adjusting the relay's position to enhance the $\Sn$-$\R$ and $\R$-$\D$ links, respectively. To gain more insight, Fig.~\ref{F:speed} shows the relay speed versus the flying time for three different trajectories shown in Fig.~\ref{F:TrajectoryUpdatesV50T40FixPowerEqualPower}. It is observed that at the converged trajectory,  the relay employs a binary speed, i.e., it remains stationary for certain duration when it reaches $\Sn$ and $\D$ and moves at the maximum speed otherwise.

\begin{figure}
\centering
\includegraphics[scale=0.48]{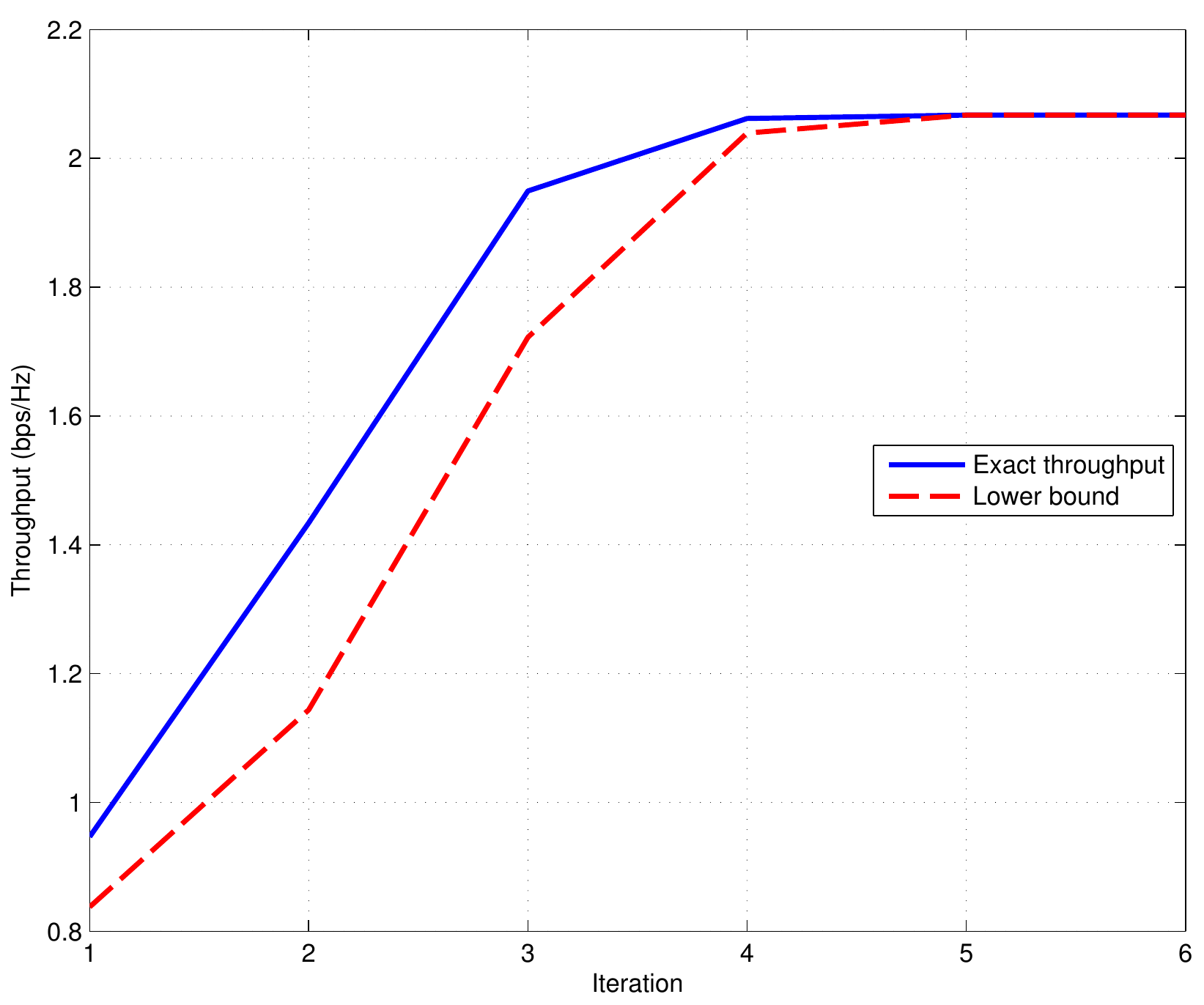}
\caption{Convergence of Algorithm~\ref{Algo:SCA}.}\label{F:convergence}
\end{figure}

In Fig.~\ref{F:convergence}, both the exact throughput and that based on the lower bound in Lemma~\ref{lemma:approx} are plotted versus the iteration number of Algorithm~\ref{Algo:SCA}. Comparing the converged throughput versus the initial throughput in Fig.~\ref{F:convergence}, it is shown that the trajectory optimization significantly improves the mobile relaying system throughput, even with constant source/relay transmit power. It is also observed that Algorithm~\ref{Algo:SCA} is quite efficient since it converges in just a few iterations. Besides, this figure shows that Lemma~\ref{lemma:approx} provides a reasonable throughput lower bound for trajectory optimization.

\subsection{Joint Power and Trajectory Optimization}
Last, we consider the mobile relaying system where the power allocation and the relay trajectory are jointly optimized for throughput maximization.  We consider the setup without pre-specified initial or final relay locations, for which the jointly optimal power allocation and relay trajectory design has been obtained in Section~\ref{sec:noRestr}. 
Besides static relaying, we also consider another benchmark scheme called {\it data ferrying}, where the carrier (e.g., the UAV) first loads the data from $\Sn$ when it is within some pre-determined range $d_1$ from $\Sn$, travels towards $\D$ without any concurrent data reception/transmission, and then unloads the data to $\D$ when it is within range $d_2$ from $\D$. For the numerical results shown below, we set $d_1=d_2=100$m.

In Fig.~\ref{F:ThroughputVSTNoInitialFinalConstraint}, the end-to-end throughput achieved by the various schemes is plotted versus the duration $T$. It is first observed that for the mobile relaying scheme, the iterative algorithm proposed in Section~\ref{sec:iterative}, which is applicable for the more general setup with initial/final relay location constraints, achieves almost identical performance as the theoretically optimal solution in Section~\ref{sec:noRestr}. Furthermore, it is observed that the optimized mobile relaying schemes significantly outperform the conventional static relaying technique. On the other hand, the data ferrying scheme performs even worse than static relaying for small $T$, which is expected since in this case, the carrier's traveling time from $\Sn$ to $\D$ is quite significant and hence only limited time is available for data loading/unloading. When $T$ gets sufficiently large so that the UAV's traveling time is negligible, data ferrying approaches to mobile relaying since in this case, both schemes essentially concentrate most of the power to time slots with the best link qualities, i.e., when the UAV is near to $\Sn$ or $\D$. 

\begin{figure}
\centering
\includegraphics[scale=0.6]{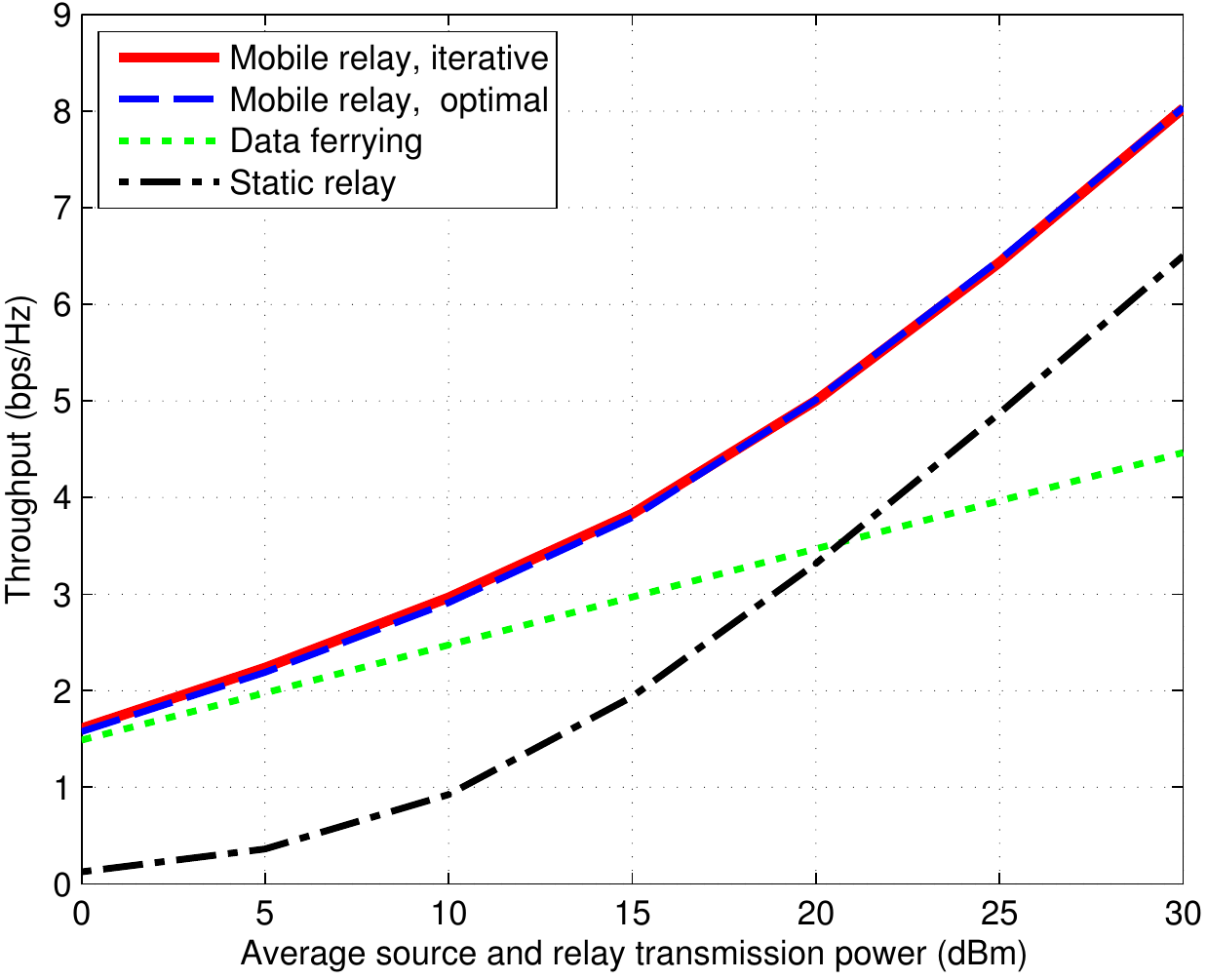}
\caption{Throughput versus average source/relay power $\bar P$.}\label{F:ThroughputVSPower}
\end{figure}

In Fig.~\ref{F:ThroughputVSPower}, the throughput is plotted against the source/relay's average power $\bar P\triangleq \bar P_s=\bar P_r$ by assuming $T=100$s. It is observed that data ferrying outperforms static relaying only in power-limited regime with small $\bar P$, whereas it performs even worse than static relaying for large $\bar P$. On the other hand, the proposed mobile relaying achieves higher throughput than both static relaying and data ferrying in all power regime.

\begin{figure}
\centering
\includegraphics[scale=0.6]{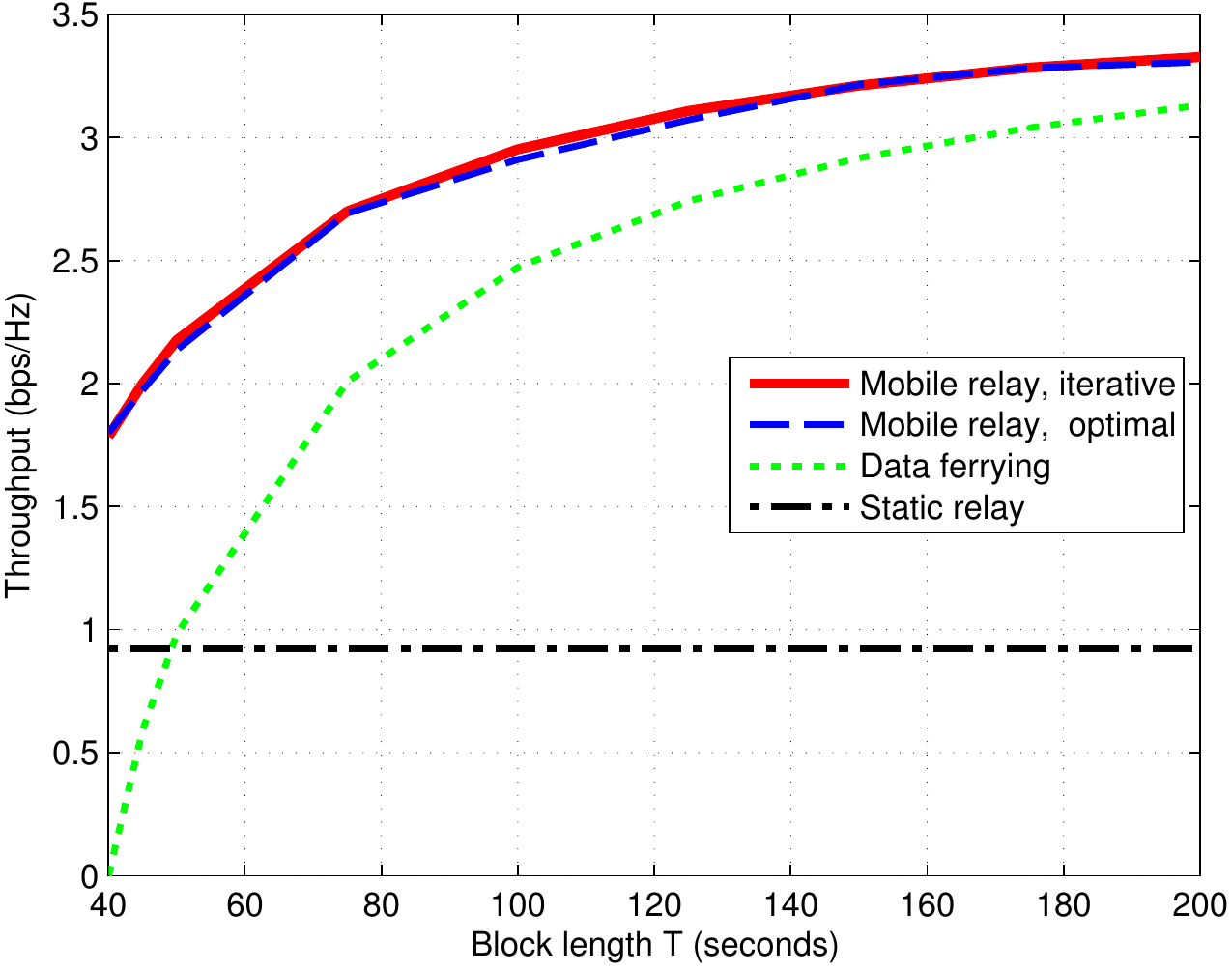}
\caption{Throughput for mobile relaying with jointly optimized power allocation and trajectory versus static relaying and data ferrying.}\label{F:ThroughputVSTNoInitialFinalConstraint}
\end{figure}

\section{Conclusions and Future Work}\label{sec:conclusion}
This paper studies a new mobile relaying technique with high-mobility relays. By exploiting the controllable channel variation induced by relay mobility, the end-to-end throughput is maximized via optimizing both the relay trajectory as well as the source/relay power allocation.  With fixed relay trajectory, it is shown that the optimal source/relay power allocation follows a staircase water filling structure with non-increasing and non-decreasing water levels at the source and relay, respectively. Besides, with given power allocation, the relay trajectory can be optimized via successive convex optimization.  Based on these results, an iterative algorithm is proposed to jointly optimize the power allocation and relay trajectory in an alternating manner. Furthermore, for the special case with free initial and final relay locations, the jointly optimal trajectory and power allocation is analytically derived. Numerical results show that compared with the conventional static relaying, a significant throughput gain is achieved by the proposed mobile relaying design, which shows the great potential of mobile relaying for future wireless communication systems.

There are several research directions along which the developed results in this paper can be further investigated, as briefly discussed in the following.

\begin{itemize}
\item {\it Fading channels:} For UAV-enabled mobile relaying, while LoS links are expected for UAV-ground channels in most scenarios, they could also be occasionally blocked by obstacles such as terrain, buildings, or even the airframe itself \cite{657}, \cite{621}. Besides, the UAV-ground channels may also constitute a number of multi-path components due to reflection, scattering, and diffraction by mountains, ground surface, and foliage, etc. Therefore, the extension of the results in this paper to the more general fading channels is an interesting topic for future research.

\item {\it Adaptive bandwidth allocation:} In this paper, it is assumed that the bandwidth allocated for the source-relay and relay-destination links are equal. The system throughput can be further improved if bandwidth allocation is also adaptively optimized based on the relay location/channel condition, which deserves further investigation.

\item {\it Limited buffer size:} In practice, the buffer size at the mobile relay is limited. The finite buffer size may lead to quite different solution for power allocation and trajectory design in mobile relaying systems, a problem that remains open.

\item {\it Throughput-delay tradeoff:} Intuitively, there exists a general tradeoff for maximizing throughput and minimizing delay in mobile relaying systems, since larger delay tolerance offers higher degrees of freedom for mobility control of the relay, and vice versa. More research endeavor is needed to resolve such a tradeoff in mobile relaying systems.
\end{itemize}

\appendices

\section{Proof of Lemma~\ref{lemma:zeroLambda}}\label{A:zeroLambda}
We show Lemma~\ref{lemma:zeroLambda} by contradiction. Suppose, on the contrary that for the dual optimal solution $\{\lambda_n^\star\}$ there exists $2 \leq n' \leq N-1$ such that $\lambda^\star_{n'}>0$. Then this must correspond to Case 1 as discussed in Section~\ref{sec:optSol}. 
Thus, the transmission rates at $\Sn$ and $\R$ corresponding to the primal optimal solution of (P1.1) can be expressed as
\begin{align}
&R^\opt_s[n]= \left[\log_2\left(\eta\beta_n^\star \gamma_{\SR}[n] \right) \right]^+, \ n=1,\cdots, N-1, \label{eq:RsStar} \\
&R^\opt_r[n]= \left[\log_2\left(\xi \nu_n^\star \gamma_{\RD}[n] \right) \right]^+, \ n=2,\cdots, N. \label{eq:RrStar}
\end{align}
 Since both $\{\beta_n^\star\}$ and $\{\gamma_{\SR}[n]\}$ are non-increasing over $n$, it follows from \eqref{eq:RsStar} that $R^\opt_s[n]$ is non-increasing over $n$ too.  We thus have
$R_s^\opt[1] \geq R_s^\opt[2] \geq \cdots  \geq R_s^\opt[n'-1]$, which implies
\begin{align}
\sum_{n=1}^{n'-1} R_s^\opt[n] \geq (n'-1) R_s^\opt[n'-1]. \label{eq:rl2}
\end{align}
On the other hand, since both $\gamma_{\RD}[n]$ and $\nu_n^\star$ are non-decreasing over $n$, it follows from \eqref{eq:RrStar} that $R_r^\opt[n]$ is non-decreasing over $n$, or $R_r^\opt[2] \leq R_r^\opt[3] \leq \cdots  \leq R_r^\opt[n']$, which leads to
\begin{align}
\sum_{n=2}^{n'} R_r^\opt[n] \leq (n'-1) R_r^\opt[n'].\label{eq:rl3}
\end{align}
Furthermore, by applying the complementary slackness condition for primal and dual optimal solutions, the assumption  $\lambda^\star_{n'}>0$ implies that the information-causality constraint at slot $n'$ must be active, i.e.,
\begin{align}
\sum_{n=1}^{n'-1} R_s^\opt[n] =\sum_{n=2}^{n'} R_r^\opt[n]. \label{eq:rl1}
\end{align}
The relations in \eqref{eq:rl2}-\eqref{eq:rl1} lead to
\begin{align}
R_s^\opt[n'-1]\leq R_r^\opt[n'].\label{eq:rl7}
\end{align}

Now consider the slots from $n'$ to $N$. Based on the non-increasing property of $R_s^\opt[n]$, we have
\begin{align}
R_s^\opt[N-1] \leq \cdots \leq R_s^\opt[n'] < R_s^\opt[n'-1],\label{eq:rl4}
\end{align}
where 
the strict inequality is true since $\lambda_{n'}^\star>0$ implies $\beta^\star_{n'}<\beta^\star_{n'-1}$, as can be seen from \eqref{eq:betan}. Similarly, we have
\begin{align}
R_r^\opt[n'] < R_r^\opt[n'+1]\leq  \cdots \leq R_r^\opt[N]. \label{eq:rl5}
\end{align}
The relations in \eqref{eq:rl7}-\eqref{eq:rl5} jointly lead to
\begin{align}
\sum_{n=n'}^{N-1} R_s^\opt[n] < \sum_{n=n'+1}^N R_r^\opt[n].\label{eq:rl6}
\end{align}
By adding \eqref{eq:rl1} and \eqref{eq:rl6}, we have
$\sum_{n=1}^{N-1} R_s^\opt[n] < \sum_{n=2}^N R_r^\opt[n]$,
which obviously violates the information-causality constraint \eqref{eq:InfoCausalConstr} at slot $N$. Thus, the assumption $\lambda^\star_{n'}>0$ for $2\leq n' \leq N-1$ is invalid. This completes the proof of Lemma~\ref{lemma:zeroLambda}.

\section{Proof of Theorem~\ref{theo:th1}}\label{A:th1}
Based on the discussions presented in Section~\ref{sec:optSol}, for any given dual optimal solution $\{\lambda_n^\star\}_{n=2}^N$, the corresponding primal optimal solution to (P1.1) can be obtained by separately considering the first three cases given in Section~\ref{sec:optSol}. In the following, we first show how to obtain the primal optimal solution to (P1.1) for Case 2. 

As discussed in Section~\ref{sec:optSol}, for Case 2, the optimal power allocation $p_s^*[n]$ at $\Sn$ is given by the classic WF solution with full power, i.e., $p_s^\opt[n]=\pcwf_{s,n}(E_s)$, $\forall n$, and the corresponding source transmission rate is $R_s^\opt[n]=\left[\log_2( \eta \gamma_{\SR}[n])\right]^+$, $\forall n$, with $\eta$ denoting the water level. Furthermore, the optimal power and rate allocations at $\R$ can be obtained by solving (P1.1) with the the obtained $R_s^\opt[n]$, i.e.,
\begin{equation}\label{eq:case2}
\begin{aligned}
   &\underset{\{p_r[n], R_r[n]\}_{n=2}^N}{\max}   \ \sum_{n=2}^N R_r[n]  \\
  \text{s.t.} \ &  \sum_{i=2}^n R_r[i] \leq \sum_{i=1}^{n-1} R_s^\opt[i], \ \forall n, \\
  &\ R_r[n] \leq \log_2\left(1+p_r[n]\gamma_{\RD}[n]\right), \ \forall n, \\
& \sum_{n=2}^N p_r[n] \leq E_r, \ p_r[n] \geq 0, \ \forall n.
\end{aligned}
\end{equation}

To solve problem \eqref{eq:case2}, we first consider its relaxed problem by discarding the information-causality constraint from slot $2$ to slot $N-1$, i.e., by solving
\begin{equation}\label{eq:case2Relaxed}
\begin{aligned}
   &\underset{\{p_r[n], R_r[n]\}_{n=2}^N}{\max}   \ \sum_{n=2}^N R_r[n]  \\
  \text{s.t.} \ &  \sum_{n=2}^N R_r[n] \leq \sum_{n=1}^{N-1} R_s^\opt[n], \\
  &\ R_r[n] \leq \log_2\left(1+p_r[n]\gamma_{\RD}[n]\right), \ \forall n, \\
& \sum_{n=2}^N p_r[n] \leq E_r, \ p_r[n] \geq 0, \ \forall n.
\end{aligned}
\end{equation}

\begin{lemma}\label{lemma:lm1}
The optimal power allocation to problem \eqref{eq:case2Relaxed} is $p_r^\opt[n]=\pcwf_{r,n}( \hat E_r)$, with $\pcwf_{r,n}(\cdot)$ and $\hat E_r$ defined in Theorem~\ref{theo:th1}.
\end{lemma}
\begin{IEEEproof}
With the function $\Rcwf_{r}(\tilde E_r)$ for any $0\leq \tilde E_r\leq E_r$ defined in Theorem~\ref{theo:th1}, it is not difficult to see that problem \eqref{eq:case2Relaxed} is  equivalent to finding the optimal total transmission power $\tilde{E}_r$ at $\R$ via solving
\begin{equation}
\begin{aligned}
\underset{0\leq \tilde{E}_r \leq E_r}{\max} \ & \Rcwf_r(\tilde{E}_r), \quad
\text{s.t. }  \Rcwf_r(\tilde{E}_r)\leq \sum_{n=1}^{N-1} R_s^\opt[n].
\end{aligned}
\end{equation}
Using the fact that $\Rcwf_r(\tilde{E}_r)$ monotonically increases with $\tilde E_r$, the results in Lemma~\ref{lemma:lm1}  can be readily obtained.
\end{IEEEproof}

\begin{lemma}\label{lemma:lm3}
If $\gamma_{\SR}[n]$ is non-increasing and $\gamma_{\RD}[n]$ is non-decreasing over $n$, problems \eqref{eq:case2} and \eqref{eq:case2Relaxed} are equivalent.
\end{lemma}

\begin{IEEEproof}
Note that problem \eqref{eq:case2Relaxed} is a relaxation of \eqref{eq:case2}. Thus, if the optimal solution to \eqref{eq:case2Relaxed} given in Lemma~\ref{lemma:lm1} is also feasible to problem \eqref{eq:case2}, then it must also be the optimal solution to \eqref{eq:case2}, and hence the two problems are equivalent. We show this by contradiction. 

Suppose, on the contrary, that the solution given in Lemma~\ref{lemma:lm1} is not feasible to problem \eqref{eq:case2}, i.e., the information-causality constraint is violated for some slot from $2$ to $N-1$. Then let $n'$ be the smallest value in $\{2,\cdots, N-1\}$ that violates the constraint, i.e., $n'$ is the slot such that $\sum_{i=2}^{n'} R_r^\opt[i]>\sum_{i=1}^{n'-1} R_s^\opt[i]$ and  $\sum_{i=2}^{n'-1} R_r^\opt[i]\leq \sum_{i=1}^{n'-2} R_s^\opt[i]$, where $R_r^\opt[i]$ denotes the optimal transmission rate by $\R$ for problem \eqref{eq:case2Relaxed}. Then we must have $R_r^\opt[n']>R_s^\opt[n'-1]$. Furthermore, since $\gamma_{\SR}[n]$ and $\gamma_{\RD}[n]$ are non-increasing and non-decreasing over $n$, we have $R_s^\opt[n]$ and $R_r^\opt[n]$  non-increasing and non-decreasing, respectively, which gives
\begin{align}
R_s^\opt & [N-1]\leq \cdots \leq R_s^\opt[n']\leq R_s^\opt[n'-1] \notag \\
& <R_r^\opt[n']\leq R_r^\opt[n'+1]\cdots \leq R_r^\opt[N]. \label{eq:Ineq}
\end{align}
The inequality in \eqref{eq:Ineq} implies that $\sum_{i=n'+1}^N R_r^\opt[i]>\sum_{i=n'}^{N-1}R_s^\opt[i]$. Together with the assumption $\sum_{i=2}^{n'} R_r^\opt[i]>\sum_{i=1}^{n'-1} R_s^\opt[i]$, we have $\sum_{i=2}^N R_r^\opt[i] > \sum_{i=1}^{N-1} R_s^\opt[i]$, which contradicts the first constraint of problem \eqref{eq:case2Relaxed}, and hence $\{R_r^\opt[i]\}$ cannot be the optimal solution to \eqref{eq:case2Relaxed}. Thus, the assumption is invalid, or the solution  given in Lemma~\ref{lemma:lm1}  must be feasible, and hence the optimal solution  to problem \eqref{eq:case2}. This completes the proof of Lemma~\ref{lemma:lm3}.
\end{IEEEproof}

With Lemma~\ref{lemma:lm1} and Lemma~\ref{lemma:lm3}, the optimal power allocation for the case when $\Rcwf_s(E_s)\leq \Rcwf_r(E_r)$ as given in Theorem~\ref{theo:th1} is obtained. For Case 1 and Case 3 given in Section~\ref{sec:optSol}, the primal optimal power allocations can be obtained similarly, which results in the solution in Theorem~\ref{theo:th1} corresponding to $\Rcwf_s(E_s)\geq \Rcwf_r(E_r)$. The details are omitted for brevity.

 This thus completes the proof of Theorem~\ref{theo:th1}.

\section{Proof of Lemma~\ref{lemma:approx}}\label{A:approx}
To show Lemma~\ref{lemma:approx}, we first define the function $f(z)\triangleq \log_2\left(1+\frac{\gamma}{A+z}\right)$ for some constant $\gamma\geq 0$ and $A$, which can be shown to be  convex with respect to $z\geq -A$. Using the property that the first-order Taylor approximation of a convex function is a global under-estimator \cite{202}, for any given $z_0$, we have $f(z)\geq f(z_0) + f'(z_0)(z-z_0)$, $\forall z$, where $f'(z_0)=\frac{-(\log_2 e) \gamma }{(A+z_0)(A+\gamma+z_0)}$ is the derivative of $f(z)$ at point $z_0$. By letting $z_0=0$, we have the following inequality
  \begin{align}
 \log_2\left(1+ \frac{\gamma}{A+z}\right)
 \geq \log_2\left( 1+ \frac{\gamma}{A}\right) - \frac{(\log_2 e) \gamma z}{A(A+\gamma)}, \ \forall z. \label{eq:taylor}
 \end{align}

The channel capacity $R_{s,l+1}[n]$ can thus be expressed as 
\begin{align}
R_{s,l+1}[n]&=\log_2 \left(1+\frac{\gamma_s[n]}{H^2+x_{l+1}^2[n]+y_{l+1}^2[n]} \right) \\
&=\log_2 \left(1+\frac{\gamma_s[n]}{d_{\SR, l}^2[n] + \Delta} \right), \label{eq:subxy}
\end{align}
where $d_{\SR,l}[n]\triangleq \sqrt{H^2 + x_l^2[n] + y_l^2[n]}$ and $\Delta\triangleq \delta_l^2[n]+\xi_l^2[n] + 2 x_l[n] \delta_l[n] + 2 y_l[n] \xi_l[n]$. Note that in \eqref{eq:subxy}, we have used the identity $x_{l+1}[n]=x_l[n]+\delta_l[n]$ and $y_{l+1}[n]=y_l[n]+\xi_l[n]$.
 As a result, \eqref{eq:RsNext} follows  from \eqref{eq:taylor} by letting $\gamma=\gamma_s[n]$,  $A=d_{\SR, l}^2[n]$,  and $z=\Delta$, and the coefficients $a_{s,l}[n]$, $b_{s,l}[n]$, and $c_{s,l}[n]$ in \eqref{eq:RsNext} can be obtained as
\begin{equation}\label{eq:CoeffS}
\begin{aligned}
 &a_{s,l}[n] = \frac{\gamma_s[n] \log_2e}{d_{\SR, l}^2[n] \big(\gamma_s[n] + d_{\SR, l}^2[n] \big)},  \\
  & b_{s,l}[n] = 2 x_l[n] a_{s,l}[n], \ c_{s,l}[n]= 2y_l[n]a_{s,l}[n],  \forall n.
\end{aligned}
\end{equation}

Similarly, the lower bound \eqref{eq:RrNext} can be obtained, and the corresponding coefficients can be obtained as
\begin{equation}\label{eq:CoeffR}
\begin{aligned}
&a_{r,l}[n] = \frac{\gamma_r[n] \log_2e}{d_{\RD, l}^2[n] \big(\gamma_r[n] + d_{\RD, l}^2[n] \big)},  \\
  & \ b_{r,l}[n] = -2 (D-x_l[n]) a_{r,l}[n], \ c_{r,l}[n]= 2y_l[n]a_{r,l}[n], \forall n,
 \end{aligned}
 \end{equation}
 with $d_{\RD,l}[n]\triangleq \sqrt{H^2 + (D-x_l[n])^2 + y_l^2[n]}$ denoting the link distance between $\R$ and $\D$ at slot $n$.

\section{Proof of Theorem~\ref{thm:nonIncreasingxn}}\label{A:nonIncreasingxn}

Denote by $\{x^\opt[n]\}$  an optimal relay trajectory to the throughput maximization problem (P1'), and $\{\gamma_{\SR}^\opt[n]\}$ and $\{\gamma_{\RD}^\opt[n]\}$ the corresponding time-dependent channels. We construct an alternative sequence $\{\tilde {x}[n]\}$ by re-ordering the elements in $\{x^\opt[n]\}$ in {\it non-decreasing} order. It can be shown that $\{x^\opt[n]\}$ is also a feasible trajectory, i.e., it satisfies the speed constraint \eqref{eq:speedContr} (recall that $y[n]=0$, $\forall n$). Furthermore, the new time-dependent channels, denoted as  $\{\tilde{\gamma}_{\SR}[n]\}$ and $\{\tilde{\gamma}_{\RD}[n]\}$ contains identical elements as $\{\gamma_{\SR}^\opt[n]\}$ and $\{\gamma_{\RD}^\opt[n]\}$, respectively, but with different orders. Let $R^\opt (\{x[n]\})$ be the optimal value of problem (P1.1) with optimized source and relay power allocations for any fixed relay trajectory $\{x[n]\}$. We aim to show that $R^\opt (\{\tilde{x}[n]\})\geq R^\opt (\{x^\opt[n]\})$, i.e., the newly constructed trajectory $\{\tilde{x}[n]\}$ achieves no smaller throughput than $\{x^\opt[n]\}$, and thus must also be optimal. We have the following relations:
\begin{align}
R^\opt \big (\{\tilde{x}[n]\} \big) & = \min \Big \{ \Rcwf_s\big(\{\tilde x[n]\}\big), \Rcwf_r\big(\{\tilde x[n]\}\big)\Big\} \label{eq:eq1}\\
&= \min \Big \{ \Rcwf_s\big(\{x^\opt[n]\}\big), \Rcwf_r\big(\{x^\opt[n]\}\big)\Big\} \label{eq:eq2} \\
& \geq R^\opt \big(\{x^\opt[n]\} \big) \label{eq:ineq3},
 \end{align}
 where \eqref{eq:eq1} follows from Theorem~\ref{theo:th1} and the fact that $\{\tilde{x}[n]\}$ is non-decreasing over $n$,  \eqref{eq:eq2} is true since $\{\tilde x[n]\}$ has identical elements as $\{x^\opt[n]\}$, or the corresponding channels are identical except the different ordering across slots, which makes no difference to the classic WF power allocation solutions, \eqref{eq:ineq3} is true since the expression given in \eqref{eq:eq2} in fact corresponds to the optimal value of problem (P1.1) by ignoring the information-causality constraints \eqref{eq:InfoCausalConstr} up to slot $N-1$, and thus it serves as an upper bound for the optimal value of (P1.1) with the fixed trajectory $\{x^\opt[n]\}$.

 This thus completes the proof of Theorem~\ref{thm:nonIncreasingxn}.

\section{Proof of Theorem~\ref{theo:th3}}\label{A:th3}
Theorem~\ref{theo:th3} can be shown by using the fact that $\Rcwf_s\big(\{x[n]\}\big)$ and $\Rcwf_r\big(\{x[n]\} \big)$ are  element-wise {\it non-increasing} and {\it non-decreasing} functions of $\{x[n]\}$, respectively. Suppose at the optimal trajectory $\{x[n]\}$, there exists a slot $n'$ such that $0<x[n']<D$ and $v[n']\triangleq x[n'+1]-x[n']<V$. Then if the $\Sn$-$\R$ link is the bottleneck, i.e.,  $\Rcwf_s\left(\{x[n]\}\right)\leq  \Rcwf_r\big(\{x[n]\} \big)$, one may slightly reduce $x[1],\cdots x[n']$ to increase $\Rcwf_s$ (while also slightly reducing $\Rcwf_r$), yet without violating the mobility constraints \eqref{eq:mobileConstr1} and \eqref{eq:mobileConstr} or decreasing the objective value of (P2). On the other hand, if the $\R$-$\D$ link is the bottleneck, one may slightly increase $x[n'+1],\cdots x[N]$ to enhance the $\R$-$\D$ link. The process continues until $v[n']=V$. Thus, without loss of optimality to (P2), we have $v[n]=V$ if $0<x[n]<D$. For $x[n]=D$, we must have $v[n]=0$, since otherwise $\R$ may move out of the interval $[0, D]$. Similarly for $x[n]=0$, $v[n]$ should be either $0$ or $V$.

This thus completes the proof of Theorem~\ref{theo:th3}.

\bibliographystyle{IEEEtran}
\bibliography{IEEEabrv,IEEEfull}

\end{document}